\documentclass[]{article}
\usepackage{parskip}
\usepackage[margin=1.0in]{geometry}

\usepackage{graphicx}
\usepackage{caption}
\usepackage{subcaption}
\usepackage{wrapfig}
\usepackage{float}

\usepackage[table]{xcolor}
\usepackage{tikz}
\usepackage{tikz-cd}
\usepackage{physics}
\usepackage{amsmath}
\usepackage{amsthm}
\usepackage{amstext}
\usepackage{amssymb}
\usepackage{thmtools}

\usepackage{booktabs}
\usepackage{makecell, multirow}

\usepackage{pifont}% http://ctan.org/pkg/pifont
\newcommand{\xmark}{\ding{55}}%

\usepackage{enumitem}

\theoremstyle{definition}
\newtheorem{defn}{Definition}[section]
\newtheorem{exmp}{Example}[section]
\newtheorem{prop}{Proposition}[section]

\usepackage[english]{babel}

\usepackage{pdfpages}

\usepackage{dsfont}
\usepackage{textcomp}
\usepackage{listings}
\usepackage{relsize}
\usepackage{multicol}
\usepackage{hyperref}
\hypersetup{
	colorlinks=true,
	linkcolor=black,
	filecolor=black,      
	urlcolor=black,
	citecolor=blue,
}
\usepackage[capitalise]{cleveref}
\crefname{figure}{Fig.}{Fig.}
\crefname{table}{Table}{Tables}
\crefname{equation}{Eqn.}{Eqns.}
\crefname{exmp}{Example}{Examples}
\Crefname{exmp}{Ex.}{Ex.}
\crefname{algocf}{Algorithm}{Algorithms}

\crefname{lemma}{Lemma}{Lemmas}
\crefname{corollary}{Corollary}{Cor.}
\crefname{defn}{Definition}{Definitions}
\crefname{prop}{Proposition}{Propositions}

\usepackage{url}

\usepackage{pifont}

\usepackage{bbm}
\usepackage{mathtools}
\usepackage{hhline}
\usepackage{tabularx}

\newcommand{\La}{\Lambda}

\newcommand{\R}{\mathbb{R}}

\newcommand{\mcM}{\mathcal{M}}
\newcommand{\mcN}{\mathcal{N}}
\newcommand{\mcG}{\mathcal{G}}
\newcommand{\mcL}{\mathcal{L}}
\newcommand{\mcT}{\mathcal{T}}
\newcommand{\mfM}{\mathfrak{M}}

\newcommand{\fnn}{f_\textsc{nn}}

\DeclareMathOperator{\sgn}{sgn}

\usepackage[ruled,vlined]{algorithm2e}

\usepackage[acronym]{glossaries-extra}
\setabbreviationstyle[acronym]{long-short}
\newacronym{nn}{NN}{neural network}

\newacronym{EOS}{EOS}{Equations of State}
\newcommand{\eos}{\gls{EOS} }

\newtheorem{problem}{Problem}

\usepackage{authblk}

\title{Neural Network Representations of Multiphase Equations of State}
\author[1,2]{George A. Kevrekidis}
\author[1]{Daniel A. Serino}
\author[1]{Alexander Kaltenborn}
\author[1]{J. Tinka Gammel}
\author[1,2]{Joshua W. Burby}
\author[1]{Marc L. Klasky}
\affil[1]{Los Alamos National Laboratory, Los Alamos, NM, USA}
\affil[2]{Department of Applied Mathematics and Statistics, Johns Hopkins University, Baltimore, MD, USA}
\affil[3]{Department of Physics, University of Texas at Austin, Austin, TX, USA}
\date{June 28, 2024\\LA-UR-24-25494}
\begin{document}

\maketitle

\begin{abstract}
    Equations of State model relations between thermodynamic variables and are ubiquitous in scientific modelling, appearing in modern day applications ranging from Astrophysics to Climate Science. The three desired properties of a general Equation of State model are adherence to the Laws of Thermodynamics, incorporation of phase transitions, and multiscale accuracy. Analytic models that adhere to all three are hard to develop and cumbersome to work with, often resulting in sacrificing one of these elements for the sake of efficiency. In this work, two deep-learning methods are proposed that provably satisfy the first and second conditions on a large-enough region of thermodynamic variable space. The first is based on learning the generating function (thermodynamic potential) while the second is based on structure-preserving, symplectic neural networks, respectively allowing modifications near or on phase transition regions. They can be used either “from scratch” to learn a full Equation of State, or in conjunction with a pre-existing consistent model, functioning as a modification that better adheres to experimental data. We formulate the theory and provide several computational examples to justify both approaches, and highlight their advantages and shortcomings.
\end{abstract}

\section{Introduction}\label{sec:Intro}
Equations of State (EoSs) are ubiquitous in scientific modelling and have been studied extensively for the better part of the past two centuries, as they are essential in a wide variety of fields including astrophysics (e.g., \cite{astro_1, astro_2,astro_3_christensen1992solar, astro_4_universe7080257,morawski2020neural,guo2023insights}), climate research (e.g., \cite{climate_1_kirk_climate_models,climate_2_krasnopolsky,climate_3_history_edwards}), hydrodynamics (e.g., \cite{zeman2019hereos}), plasma models in nuclear fusion (e.g., \cite{gaffney2018review}), and asteroid impact modelling (e.g., \cite{thomas2006comets}), among others. In their most elementary form, EoS models consist of closed-form equations, such as the ideal gas law \cite{Clausius1855}, Van der Waals \cite{van1873over}, and those derived from statistical mechanics e.g., the Virial equation \cite{onnes1991expression,dymond1980virial}. These analytic models are usually not universally accurate across many elements or even different phases of a single element.

\subsection*{Background}

Physical properties of fluids can, and have, been measured experimentally since the dawn of modern science. These properties include densities, vapor pressures, critical point data, viscosities, solubilities, surface tensions, thermal conductivities, etc., and are collated in existing open-access and closed-source databases \cite{zhu2020generating}. However, the cost of experimental acquisition of the data is at odds with the massive phase space for both pure and mixture systems of interest. Consequently, there has been a sustained effort to establish empirical relationships characterizing the EoS, enabling the determination of relationships among state variables such as density, pressure, and temperature. Indeed, the transformative work of Van der Waals provided insights into the pressure-volume temperature relationship for both liquids and gases and the interpretation of a critical condition that limited the coexistence of these phases. Using a theoretical procedure, Van der Waals derived a simple mathematical function, a third degree polynomial, capable of fitting experimental data for simple fluids. This model has guided a century of research into finding the ultimate mathematical function that can relate the macroscopic thermodynamic properties of fluids to each other \cite{kontogeorgis2020equations,chaparro2023development}.

Over the next century, the development of a more general theory connecting molecular model parameters with macroscopic observables emerged, driven by the recognition of matter's molecular nature and advances in statistical mechanics \cite{chaparro2023development}. While these simplified models have proven useful, they are limited by various issues hindering accurate EoS characterization across the extensive phase space typically encountered \cite{Rosenberger_EOS_autodiff}. Indeed, the development of an EoS has been marked by a long and convoluted search, frequently yielding unsatisfactory outcomes, underscoring the absence of a universally accepted EoS, even for the simplest fluids \cite{chaparro2023development,kontogeorgis2020equations}. Consequently, the majority of high-pressure EoS are empirical in nature and rely on limited sets of experimental data to reproduce the relevant physics \cite{kontogeorgis2020equations}. Examples of such models include the Jones Wilkens Lee (JWL), Davis reactants, Mie-Grüneisen, and Tillotson EoSs \cite{kury1960energy,lee1968adiabatic,Davis1998EquationOS,mie1903kinetischen,gruneisen1912theorie,tillotson1962metallic}. Many of these models are based on a reference curve that can be a Hugoniot, an isentrope, or an isotherm. These forms in general allow for simple linear extrapolation and are fairly accurate when the desired properties are in close proximity to the reference curve. However, significant deviations from empirical reality arise when attempting to move away from the reference curve \cite{lozano2023analytic, mie1903kinetischen, gruneisen1912theorie, lemons1999thermodynamics}.

Current state-of-the-art methods such as SESAME, ANEOS, and HerEOS (\cite{aneos_thompson1974improvements,aneos_melosh2007hydrocode,aneos_thompson2019m,zeman2019hereos}) produce `global', multi-phase EoS models. They achieve this by representing thermodynamic variables with (often analytic) techniques across different scales and phases to interpolate among predefined grid points. This approach aims to create a `complete' thermodynamic model. However, the discrete values and their distribution might be less than ideal, often subject to interpolation errors. These errors can disrupt thermodynamic consistence and negatively affect calculations dependent on these properties \cite{zeman2019hereos}. Furthermore, efficient processing of the full range of a multi-phase table requires logarithmic sampling of the phase space, which may not provide sufficient resolution to identify phase transitions from the tabular data, also requiring additional information to be provided to the model. Nevertheless, these EoS models have been utilized for the last several decades. A general review of present challenges with state-of-the-art EoS models is available, see \cite{present_challenges_EOS}.

\subsection*{Relevant Literature}

The emergence of machine learning has given promise to a new approach for building multi-phase EoS tables for downstream applications. Indeed, recent investigations have focused on addressing the question as to whether ML methods may be designed to substitute for analytical models \cite{arce2018thermodynamic,zhu2020generating,Rosenberger_EOS_autodiff,chaparro2023development}. In this regard, Arce et al. compared the quality of fit for common EoS binary systems involving solvents and bio-diesel components at supercritical conditions, finding good agreement among the EoS, ML model, and experimental data \cite{arce2018thermodynamic}. In a separate study, researchers explored the potential of using ML to correlate extensive datasets of physical properties. These datasets were produced using cutting-edge analytical EoS, specifically the Statistical Associating Theory variable range Mie equation (SAFT-VR-Mie) EoS for pure fluids \cite{zhu2020generating}. Further research has been conducted on utilizing ML to develop EoSs that \textit{exactly} adhere to energy conservation principles. This research combines these EoS with the robust, expressive power of black- or grey-box ML methods \cite{Rosenberger_EOS_autodiff, chaparro2023development}. In these, the general approach centers around estimating a smooth energy potential (e.g., Helmholtz free energy), and differentiating it, through automatic differentiation, to generate the remaining relevant thermodynamic quantities. The advantage of this approach is that the model is exactly consistent with thermodynamics. The model has been tested for the van der Waals, Lennard-Jones fluids, as well as Mie fluids \cite{Rosenberger_EOS_autodiff,chaparro2023development}. 

The latter approach is outlined in \cref{sec:graph_approach}. However, near phase transitions one may encounter difficulty using a smooth neural network since the approximation will not be consistent (\cref{ex:graph_corr}). Using non-smooth activation functions, such as ReLU, would give a network the expressivity needed to model the transitions; however, it is exceedingly hard to control their `location' in phase space numerically, so the right number and topology of phase transitions using such a model would not be guaranteed, at least without further structural modifications of the architectures. Initial attempts to address this issue have been made using multiple networks that are used for different phases (or scales) of an EoS \cite{Mentzer_surrogate_eos}. While this allows for a model to perform well in the multiscale setting that is natural to the problem at hand, one cannot guarantee that `gluing' these models around phase transitions \textit{is} thermodynamically consistent. This is a fundamental limitation to global EoS modelling with neural networks, which are known to not perform well in wide, multiscale settings \cite{elizar2022review}.

In summary, it is technically difficult to develop neural EoS models that simultaneously combine thermodynamic consistence, multi-scale accuracy, and robust handling of phase transitions, which are all desirable features of a general framework.

\subsection*{Structure and Contributions}

In this work, we develop two general methodologies that can represent Equations of State numerically, \textit{exactly} satisfying conservation of energy and simultaneously incorporating phase transitions.

We view this problem through a geometric framework, where the symplectic character of thermodynamic manifolds (which, by definition, satisfy conservation of energy) is used to analyze our approach (\textbf{\cref{sec:problem_statement}}). Based on this, we outline two EoS models that can generate thermodynamic data consistent with conservation of energy (\textbf{\cref{sec:methods}}): the additive (i.e., residual) correction (AGC) can do so \textit{in the neighborhood of} phase transitions without changing their support, while the symplectic correction (SC) can transform regions where the phase transition \textit{itself} occurs. We demonstrate applications of both methods through various computational examples and discuss their advantages and shortcomings. Using the output of these methods, we demonstrate that they can both be successfully utilized for further hydrodynamic computations (\textbf{\cref{sec:hydrocode}}), before concluding with several remarks (\textbf{\cref{sec:discussion}}). We use the appendix to supply the reader with additional mathematical background and to expand on properties of the symplectic network architectures used.

Our novel contributions in this work are:
\begin{itemize}[label=\ding{70}]
    \item We develop two neural EoS models (the AGC and SC) that are thermodynamically consistent \textit{and} incorporate phase transitions. The SC works by smoothly (and symplectically) deforming an already-consistent model that may include phase transitions. This bypasses the complexity of incorporating discontinuities within network architectures, allowing us to make use of traditional smooth optimization algorithms.
    \item We provide a number of numerical examples using generated tabular SESAME data. We demonstrate that our predictions can be used to generate new tables on which hydrodynamic simulations can be performed.
\end{itemize} 

\subsection*{Code and Data Accessibility}
The accompanying code will be available after publication upon reasonable request to the authors. Data related to SESAME tables can be requested through Los Alamos National Laboratory at \url{https://www.lanl.gov/org/ddste/aldsc/theoretical/physics-chemistry-materials/sesame-database.php}

\section{Problem Statement}\label{sec:problem_statement}

\textit{Note: To enhance clarity, throughout this work we will use temperature ($T$) and volume ($V$) as independent variables, while treating entropy ($S$) and pressure ($P$) as dependent variables, obtained as the partial derivatives of the Helmholtz free energy ($A$). Reformulating for alternative choices independent and dependent variables is straightforward \cite{mattsson2016short}.}

\subsection*{Setup}
The EoS modelling problem can be appropriately framed in the context of symplectic geometry. Two relevant definitions are those of a symplectic manifold and Lagrangian submanifold:

\begin{defn}
    A symplectic manifold $\mcM$ is an even-dimensional manifold equipped with a closed, non-degenerate differential two-form $\omega$, called the symplectic formn. We often denote this pairing as $(\mcM,\omega)$
\end{defn}

\begin{defn}
    A Lagrangian submanifold of a symplectic manifold $(\mcM,\omega)$ with dimension $2n$, is an $n$-dimensional submanifold $\La\subset\mcM$ on which the symplectic form vanishes, i.e. $\omega\vert_\La=0$.
\end{defn}

We define thermodynamic phase space as $\Phi\simeq\R^4$ parametrized by $(T,V,S,P)$. This space, equipped with the differential form $\omega=-dT\wedge dS-dV\wedge dP$ has symplectic structure. When viewing $S$ and $P$ as functions of the independent variables $T,V$, the condition that $\omega$ vanishes is equivalent to satisfying the following partial differential equation:
\begin{align*}
    -\pdv{S}{V}+\pdv{P}{T}=0,
\end{align*}
which can be easily seen to be true by equality of mixed partials if there exists a $C^2$-scalar function $A(T,V)$ such that $$S(T,V)=-\pdv{A(T,V)}{T}\qc P(T,V)=-\pdv{A(T,V)}{V}.$$

Then, for any particular choice of thermodynamic potential $A(T,V)$, the set of points that satisfy
\begin{align}\label{eqn:A_generator}
    \qty{(T,V,S,P)\in\Phi:S=-\pdv{A}{T},P=-\pdv{A}{V}}
\end{align}
will lie on a Lagrangian submanifold $\La\subset\mcM$.

Thus, for a given chemical compound or element, its corresponding EoS is equivalent to a two-dimensional graph (and submanifold) $\La$ in $\Phi$, which satisfies `conservation of energy'. We also refer to this property as `\textbf{energy-consistence}'. If an $2$-dimensional submanifold is not generated by some scalar function $A(T,V)$ then it does not satisfy the first law of thermodynamics.

We briefly expand on this formalism in \cref{ap:Geometric_structure}, while a more rigorous recent account can be found in \cite{aragon2022symplectic_thermo}. 

For EoS that span multiple phases, it is known that $A(T,V)$ is smooth within a single phase but may have kinks on the boundaries between phases. When phase transitions occur $A(T,V)$ may only have well-defined first- or second-order derivatives but none of higher order. In that case, these regions respectively manifest as jumps or kinks in the corresponding Lagrangian submanifold $\La$ (for details, see \cref{sec:phase_transitions}).

\subsection*{Statement}
Within this framework, we consider the following problem:

%\subsection*{Problem Statement}
{\problem \label{sec:prob_statement}
Given `input' samples of an EoS, lying on a Lagrangian submanifold $\La\subset\Phi$ and `target' samples of a distinct EoS, lying on a different Lagrangian submanifold $\La'\subset\Phi$, we are interested in learning, i.e., estimating, a correction map
\begin{align}
	f:\La\to\Phi\qq{such that} f(\La)={\La}'
\end{align}
Let our estimate correction map be $\hat{f}$. We require $\hat{f}$ 
to be energy-consistent, that is $\hat{\La}'\doteq\hat f(\La)$ must also be a Lagrangian submanifold of $\Phi$ that satisfies conservation of energy, and approximates $\La'$}

Satisfying conservation of energy is not inherently guaranteed by straightforward regression techniques between sampled submanifolds, but can still be achieved by imposing further structure on $\hat f$. However, energy-consistence is not overly restrictive, as it is satisfied by models that might be `very far from reality'. For example, modelling the energy as a constant with vanishing partial derivatives is thermodynamically consistent, but not realistic for any generic EoS.

From a physical point of view, $\La$ can be thought of as a model we want to `fit' to observations that are the samples of $\La'$. We assume that the Lagrangian submanifolds are parametrized as
\begin{align}
	(T,V,S,P)\in\La\subset\Phi\qc (T',V',S',P')\in\La'\subset\Phi
\end{align}
and that there exist functions
\begin{equation}
	q:(T,V)\mapsto Q\qc q':(T',V')\mapsto Q'
\end{equation}
that generate $\La$, $\La'$ respectively through \cref{eqn:graph_condition}
\begin{equation}\label{eqn:graph_condition}
    \begin{split}
        S\doteq s(T,V)=-\pdv{A(T,V)}{T}\\
        P\doteq p(T,V)=-\pdv{A(T,V)}{V}
    \end{split}
\end{equation}
where $q,q'$ approximate the corresponding free energy functions $A,A'$ up to an additive constant. We specify the sense in which we expect $\hat\La'\approx\La'$ later on.

\section{Methods}\label{sec:methods}
We describe two qualitatively different approaches that attempt to solve the EoS approximation problem outlined in \cref{sec:prob_statement}. Briefly, the \textit{Graph} and \textit{Additive Graph} corrections (GC, AGC) center around estimating and subsequently differentiating an `energy function' $Q$, while the \textit{Symplectic Correction} (SC) aims at manipulating a prescribed EoS in a manner that preserves conservation of energy.

\subsection{The (Additive) Graph Correction}\label{sec:graph_approach}
\subsubsection*{Formulation}
A simple way obtaining a model that exactly conserves energy comes from exploiting the structure of \cref{eqn:graph_condition} using the automatic differentiation capabilities of neural networks. We first observe that the $(T,V)$ coordinates of $\La$ can be identified with the coordinates $(T',V')$ of $\La'$ since the Lagrangian submanifold can be seen as the 2-dimensional graph of a function with independent variables $(T,V)$.

Then, in particular, if $q$ is parameterized by a differentiable (fully-connected feed forward) neural network such that
\begin{align*}
	q_{nn}:\R^2\to \R \qc (T,V)\mapsto \hat Q'
\end{align*}
obtaining its exact partial derivatives through automatic differentiation yields an energy-consistent model
\begin{equation}\label{eqn:graph_corr}
    \begin{split}
        \hat f_G:(T,V,S,P)&\mapsto(T,V,\hat S', \hat P')\\
	   \hat S'=-\pdv{q_{nn}}{T}&\qc\hat P'=-\pdv{q_{nn}}{V}
    \end{split}
\end{equation}

This method is effectively proposed in \cite{Rosenberger_EOS_autodiff}.

Note that such a method does not require information about the dependent variables $(S,P)$ on the input manifold $\La$, which may be advantageous if these are not energy-consistent when given as input. However, in the case where we have consistent (full) samples of both an input and target submanifold, we may use the same formulation to produce a more `efficient' \textit{additive} correction as follows:

Under the assumption that we can decompose the energy functional of the target submanifold as
\begin{equation}
	q'=q+q^+,
\end{equation}
where we only alter the energy functional of the input space $q$ by an additive correction $q^+$ we have, by linearity of the partial derivative, that
\begin{align}
	S'=-\pdv{q'}{T}=-\pdv{q}{T}-\pdv{q^+}{T}=S-\pdv{q^+}{T}\qc P'=-\pdv{q'}{V}=-\pdv{q}{V}-\pdv{q^+}{V}=P-\pdv{q^+}{V}
\end{align}
Thus, estimating \textit{only} this additive term with a neural network $q_\text{nn}^+\approx q^+$ allows us to formulate the following model
\begin{equation}\label{eqn:add_graph_corr}
    \begin{split}
        \hat f_G^+:(T,V,S,P)\mapsto&(T,V,\hat S', \hat P')\\
	\hat S'= S - \pdv{q_\text{nn}^+}{T}\qc\hat{P}'&=P-\pdv{q_\text{nn}^+}{V}    
    \end{split}
\end{equation}

This model makes full use of the `input' manifold information and may generally require less training, i.e., may converge faster, if the input manifold $\La$ is already a good approximation to $\La'$. Furthermore, if $\hat{f}_G$ of the GC is parametrized by a smooth neural network, it will fail when attempting to estimate the energy functional \textit{near} a phase transition, and the resulting dependent variables $(S,P)$ may be severely distorted (as can be seen in the example of \cref{fig:Graph_Approach_Simulations}). In contrast, $f_G^+$ of the AGC uses pre-existing information about the energy functional of the input space implicitly, through the ($S,P$) values of the input submanifold. Thus, as long as $q$ does not need to be `corrected' exactly on the points where a phase transition occurs, this model is capable of producing faithful approximations of the target EoS submanifold. Note that both methods fail when the phase transition region itself needs to be corrected. We revisit the comparison between the methods in \cref{sec:discussion}.

\subsubsection*{Computational Example}
\begin{exmp}[Perturbed SESAME tables I]\label{ex:graph_corr}
We demonstrate the behavior of the two methods (GC, AGC) near a phase transition by applying them to the following case study: We generate a multiphase EoS table for lead (Pb) as our ground truth target sample of $\La'$, and artificially perturb it near a phase transition region, to produce a second, perturbed table, as our input samples of $\La$ (for details on the data sets and architectures used, see \cref{app:data_architectures}).

\begin{figure}[H]
  \begin{subfigure}[b]{0.33\textwidth}
    \includegraphics[width=\textwidth]{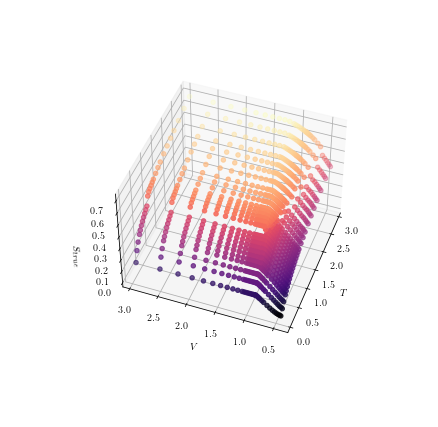}
    \caption{Ground Truth}
  \end{subfigure}
  \begin{subfigure}[b]{0.33\textwidth}
    \includegraphics[width=\textwidth]{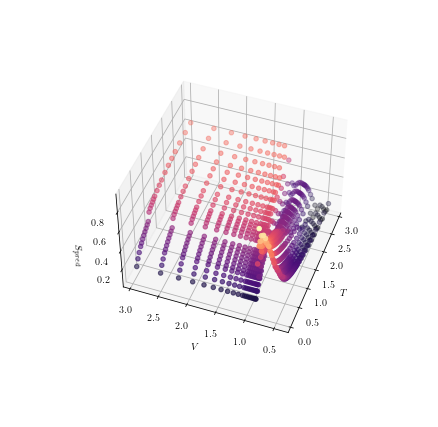}
    \caption{Graph correction $\hat{f}_G$}
  \end{subfigure}
  \begin{subfigure}[b]{0.33\textwidth}
    \includegraphics[width=\textwidth]{Paper_Images_Final/example_graph_training_TVS_true.png}
    \caption{Additive graph correction $\hat{f}_G^+$}
  \end{subfigure}
  \caption{Ground truth entropy submanifold $(T,V,S)$ compared to model predictions for the graph correction models $\hat{f}_G$ and $\hat{f}_G^+$ of section \cref{sec:graph_approach}. The color corresponds to the entropy value ($z$-axis).}
\label{fig:Graph_Approach_Simulations}
\end{figure}

We train two models $\hat{f}_G$ and $\hat{f}_G^+$ to estimate the point-wise correction between the two manifolds. For both, we minimize the supervised objective
\begin{align}
    \mcL=\frac{1}{N}\sum_{i}^N\qty(\norm{(\hat{S}'_i,\hat{P}'_i)-(S'_i,P'_i)}_2^2)
\end{align}
to achieve comparable training losses (on the order of $10^{-2}$). However, upon plotting the predicted entropy variables (\cref{fig:Graph_Approach_Simulations}) we observe that $\hat{f}_G$ has a qualitatively wrong feature, a large jump, stemming from the inability of the smooth model to capture phase transitions. This is not observed in the case of the additive correction. Fundamentally, GC fails due to the smoothness of the approximating networks (fully connected with $\tanh$ activation functions), which cannot appropriately approximate the underlying free energy. While the AGC is smooth as well, it `retains' the phase structure of the input submanifold and thus is capable of approximating the target EoS.

\end{exmp}

\subsection{The Symplectic Correction}\label{sec:symp_approach}
\subsubsection*{Formulation}
An alternative approach to the correction problem that still guarantees energy conservation makes use of symplectic diffeomorphisms, or `symplectomorphisms', of the ambient thermodynamic space $\Phi$. Informally, these are a class structure-preserving diffeomorphisms between symplectic manifolds that preserve conservation laws of their input space. For details, refer to \cref{app:Differential_Geometry}.

Universal parametrizations of symplectomorphisms via neural networks 
include the SympNet~\cite{sympnet} and HenonNet~\cite{henonnet}, which 
are based on a characterization developed in \cite{Turaev_2002}. In our work, we specifically modify the implementation proposed in \cite{henonnet} such that it is $L$-bi-$Lipschitz$. Universality guarantees that a large enough architecture will be expressive enough to approximate the desired correction map, while Lipschitz control offers some stability during training with larger learning rates. We expand on the architecture and its properties in \cref{app:symplectic_networks}.

If $\varphi:\Phi\to\Phi$ is a symplectomorphism, then the image $\varphi(\Lambda)$ of any Lagrangian submanifold is \textit{also} Lagrangian. This motivates us to introduce a symplectic neural network $\varphi_\text{nn}:\Phi\to\Phi$ whose weights are chosen to ensure $\varphi_\text{nn}(\Lambda)\approx \Lambda'$.

The desired correction map $\hat{f}_s$ is given by the restriction of $\varphi_\text{nn}$ to the input submanifold $\La$. 

In this approach, we may choose to identify the independent variables ($(T,V)$ with $(T',V')$) of the input and target submanifolds, similarly to \cref{sec:graph_approach}. However, under this identification of the independent variables, phase transitions are `locked in place'. This means that the approximation map will attempt to smooth out kinks at some location and create kinks in another if they do not line up between $\La$ and $\La'$ a priori. (For an illustrative example, see \cref{exmp:toy} in the appendix.)

Instead, we may choose to consider more general \textit{manifold} losses between $\La$ and $\La'$ where no pointwise correspondence is specified. These are metrics that consider the manifold as a whole object (e.g., Hausdorff distance between two Euclidean-embedded submanifolds, sampled as point clouds). We expand on the types of losses considered here in \cref{app:symplectic_networks}. Importantly, considering manifold distances does not fix the location of the phase transitions, only their \textit{type}: 

Since $\phi_\text{nn}$ is a diffeomorphism, it must preserve the regularity of its pre-image, and so if $\La$ has a phase transition, $\La'$ will also have \textit{the same type} of phase transition. This is desirable, since it guarantees that we have not altered the qualitative structure of the element we are studying when estimating the correction map.  Thus, by using symplectomorphisms on ambient space, we guarantee that the correction is energy consistent \textit{and} gain additional freedom in our ability to \textit{move} and \textit{model} phase transitions.

Importantly, this also gives rise to the idea of \textbf{`templating'}, where a general template (that is qualitatively correct in some sense) is prescribed and molded, i.e., fitted,to a particular data set,  not necessarily using ML methods. This framework is far more general than its use in the specific work (\cref{sec:Templating}).

\subsubsection*{Computational Examples}
\begin{exmp}[Perturbed SESAME tables II]\label{ex:symp_corr} In the identical setting of \cref{ex:graph_corr}, we estimate correction maps $\hat{f}_s$ by minimizing both the $L_2$ (pointwise) and manifold losses. These are given by \cref{eqn:L2,eqn:LH_star} respectively. In both cases, depicted in \cref{fig:Symplectic_Approach_Simulations}, the discrepancy between the estimate and target submanifolds is small after successful training (\cref{tbl:computation Results}).

\begin{figure}[!h]
  \begin{subfigure}[b]{0.495\textwidth}
    \includegraphics[width=\textwidth]{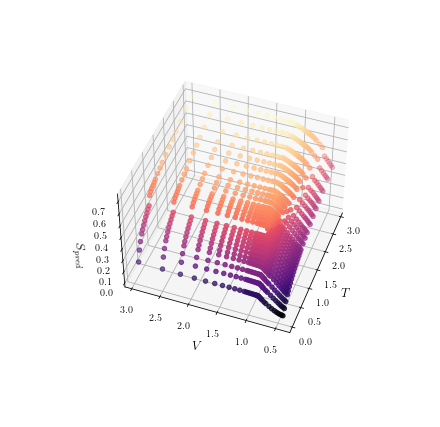}
    \caption{$L^2$ symplectic correction}
  \end{subfigure}
  \begin{subfigure}[b]{0.495\textwidth}
    \includegraphics[width=\textwidth]{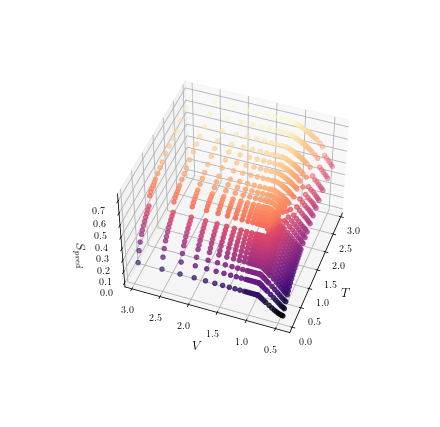}
    \caption{Manifold-loss symplectic correction}
  \end{subfigure}
  \caption{Entropy submanifold $(T,V,S)$ for a symplectic network \cref{sec:symp_approach} optimized under $L^2$ (left) and a permutation invariant Hausdorff-like loss (right). The color corresponds to the entropy value ($z$-axis).}
\label{fig:Symplectic_Approach_Simulations}
\end{figure}

We include this example to demonstrate that the symplectic correction also overcomes the phase transition issue demonstrated in \cref{ex:graph_corr}, and can be optimized through the two different types of objectives that we consider. However, a small residual and consistent plots may be insufficient to assess the quality of a fit. We consider additional validation steps through hydrodynamic simulations in \cref{sec:hydrocode}.
\end{exmp}

\begin{figure}[h!]
  \begin{subfigure}[b]{0.495\textwidth}
    \includegraphics[clip, trim={2cm 3.2cm 2cm 6cm}, width=\textwidth]{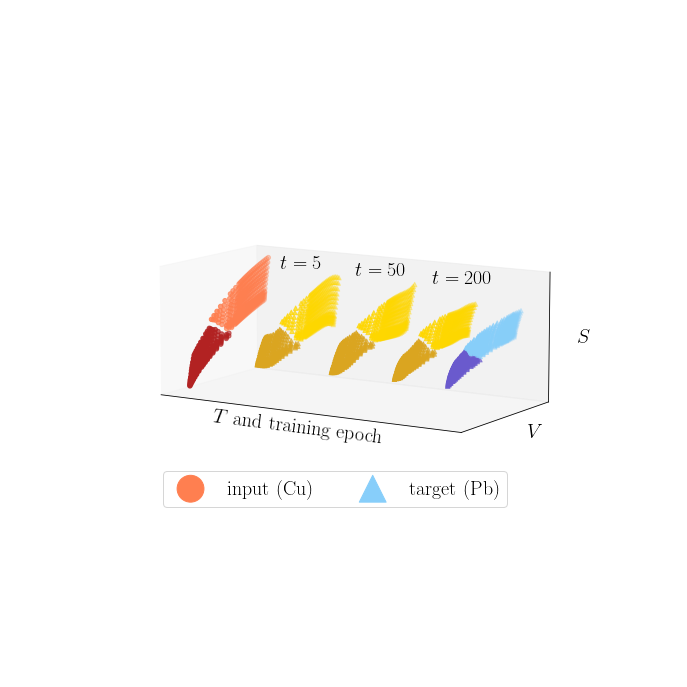}
    \caption{Entropy submanifold}
  \end{subfigure}
  \begin{subfigure}[b]{0.495\textwidth}
    \includegraphics[clip, trim={2cm 2cm 2cm 6cm}, width=\textwidth]{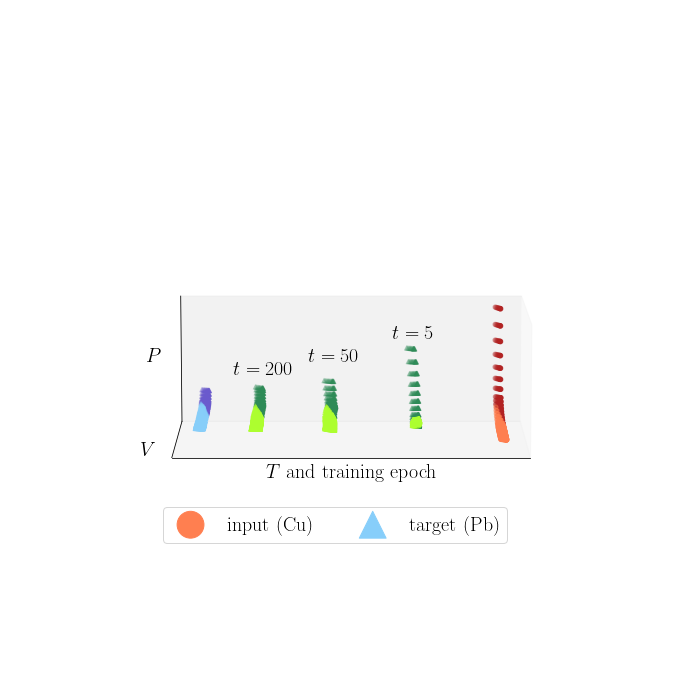}
    \caption{Pressure submanifold}
  \end{subfigure}
  \caption{Training trajectory for the entropy (left) and pressure (right) submanifolds as a symplectic network learns the map from Cu (red) to Pb (blue). The $x$-axis is translated at each training time to produce the figures: in reality the $(T,V)$ support of the submanifolds is identical, with the differences observed on the dependent variables.}
\label{fig:Cu_to_Pb_example}
\end{figure}

\begin{exmp}[Mapping Cu to Pb]\label{ex:Cu_to_Pb}
As in example \cref{ex:graph_corr}, we generate SESAME tables for lead and copper (Cu). We consider the part of the table supported on a grid of $(T,V)$ in the same range for both elements, which also includes a solid-to-liquid phase transition. We then estimate a correction $\hat{f}_s$ to map one table onto the other, noting that the dependent variables $(S,P)$ cover different ranges of values for each element, as presented in \cref{fig:Cu_to_Pb_example}.

While this example may seem extreme at first glance, it primarily serves to show that the symplectic correction can be implemented between manifolds that have a `large' distance between them, i.e., they are not within a small perturbation of each other, in example \cref{ex:graph_corr}.

We note that, if the input sesame table is thermodynamically inconsistent (which is generically the case due to interpolation errors) the output of the symplectic network will \textit{also} be inconsistent. This does not prohibit us from solving the corresponding optimization problem.
\end{exmp}

\begin{exmp}[Analytic Correction]\label{ex:analytic}

\begin{figure}[h!]
    \centering
    \includegraphics[width=0.8\textwidth]{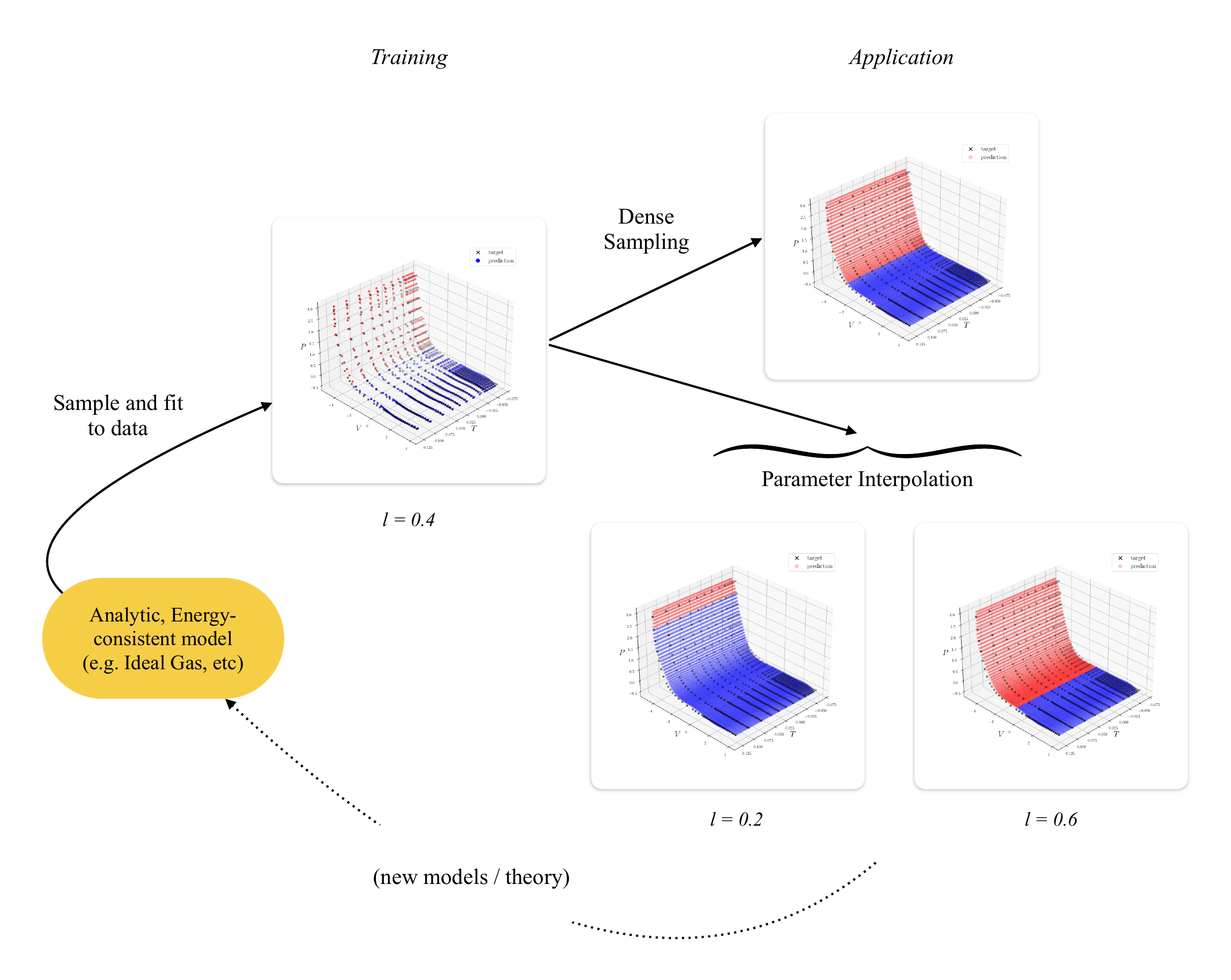}
    \caption{Representation of the workflow using an analytic template (\cref{eqn:analytic_transition}) to fit a region of SESAME table data (Cu) using a symplectic network. After fitting, we can sample arbitrary points on the fitted submanifold, thus producing dense samples. The parameters can also be modified to better fit data, (in this case different values of $l$ are shown, representing the location of the phase transition).} 
    \label{fig:analytic_workflow}
\end{figure}

Finally, we construct a simple, \textit{exact} EoS using the Ideal Gas and Sackur-Tetrode (\cite{schroeder1999introduction,sackur_tetrode_100_years}) equations. This yields the following expression for the Helmholtz free energy:

\begin{equation}
    A(T,V)\doteq\frac{3 k_B T}{2}-k_B T \qty(\ln(2 \sqrt{2} \pi ^{3/2} V \qty(\frac{k_B m T}{h^2})^{3/2})+\frac{5}{2})
\end{equation}

where $k_B$ is the Boltzmann constant, $h$ is the Planck Constant, and $m$ is the mass of a single particle. To demonstrate how this may be used as an editable template, we consider a smooth, piece-wise addition which acts as a first-order phase transition for the modelled material (\cref{eqn:analytic_transition}). (In a similar manner, one may add an arbitrary number of phase transitions of different types and at different initial locations). We consider the initial location ($l$) and scale ($s$) of the phase transition as two positive, real-valued parameters which can later be changed:

\begin{align}\label{eqn:analytic_transition}
    A_{(l,s)}(T,V)=
    \begin{cases}
    \frac{3 k_B T}{2}-k_B T \qty(\ln(2 \sqrt{2} \pi ^{3/2} V \qty(\frac{k_B m T}{h^2})^{3/2})+\frac{5}{2}),&V<l\\
    \frac{3 k_B T}{2}-k_B T \qty(\ln(2 \sqrt{2} \pi ^{3/2} V \qty(\frac{k_B m T}{h^2})^{3/2})+\frac{5}{2})-s(V-l),&V\geq l\\
    \end{cases}
\end{align}
This free energy will produce a phase transition along $V=l$ of magnitude $s$, which will result in a jump for pressure at the corresponding location.

Using this equation, we generate a sample of the analytic thermodynamic submanifold using the same $(T,V)$ values that appear in a region of the Cu SESAME table. We then proceed to estimate the correction between the analytic model and the tabular data using a symplectic network.

After convergence, we can use the analytic model to (a) generate denser samples of the tables, and (b) a posteriori modify parameters of the input template to sample an EoS in the region of the target, in an interpretable manner. Both of these capabilities can be broadly exploited in downstream tasks (\cref{fig:analytic_workflow}). Note that, assuming the network architecture used is differentiable, the output can also be differentiated with respect to parameters, thus making the method amenable to (a posteriori) parameter inference using gradient descent methods \cite{GUAN2022114217}.

\end{exmp}

\section{Closing the Loop: Hydrocode Computations}\label{sec:hydrocode}

\begin{figure}[h!]
    \centering
    \includegraphics[width=0.6\textwidth, trim={0 4cm 20cm 4cm},clip]{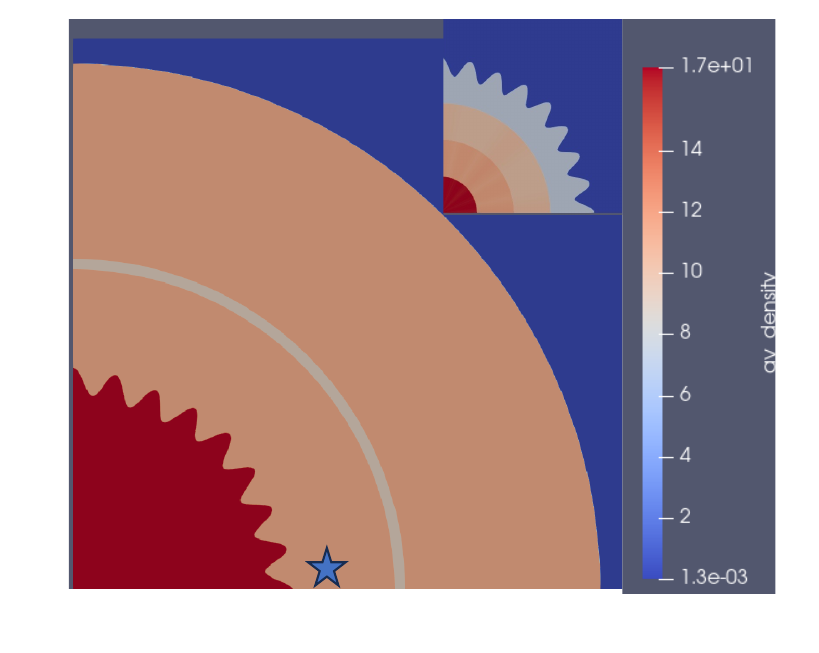}
    \caption{Initial System  Pb--Cu system with Tantalum Driving Shells and location of Lagrangian Tracer .} 
    \label{fig:InitialSystem}
\end{figure}

EoS models have numerous applications, one of which includes hydrodynamic simulations that track trajectories in thermodynamic phase space under specified geometric, boundary, and initial conditions. In this context, we explore the efficacy of our ML-generated EoS tables in facilitating hydrodynamic simulations where the thermodynamic trajectory evolves though multiple phase transitions. Our focus extends to simulations where hydrodynamic responses are intricately tied to the properties of the EoS. To this end, we delve into a specific hydrodynamic scenario characterized by the emergence of a Richtmyer-Meshkov instability (RMI), triggered by the interaction of a shock wave with a non-uniform interface between two materials of differing properties, such as density variations. An illustrative example of this phenomenon is depicted in \cref{fig:InitialSystem}, showcasing an explosive scenario where a high-density material, Tantalum (Ta), encapsulated within four concentric shells each imparted with an initial outward velocity, propels into a composite medium of lead and copper.

\begin{figure}[h!]
    \centering
    \begin{subfigure}[b]{0.43\textwidth}
    \includegraphics[width=1.1\textwidth]{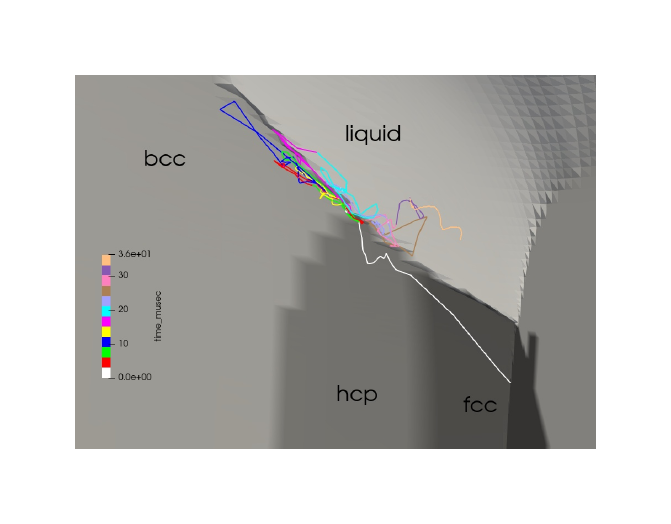}
    \caption{}
    \label{fig:Thermodynamic_Traj}    
    \end{subfigure}
    \begin{subfigure}[b]{0.43\textwidth}
        \includegraphics[width=1.1\textwidth]{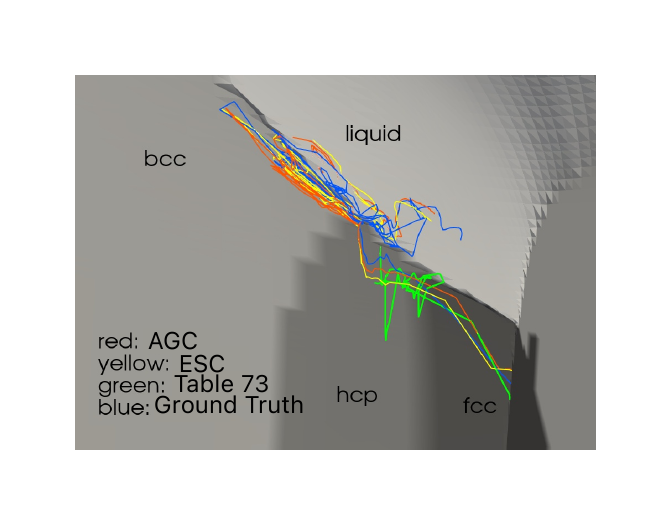}
    \caption{}
    \label{fig:ReconstructPaths}
    \end{subfigure}
    \caption{\textbf{(a)}, Thermodynamic trajectory of Lagrangian tracer in Lead. \textbf{(b)} Comparisons of the thermodynamic trajectory of the Lagrangian tracer using the baseline, initially perturbed Pb EoS table, and the reconstructed EoS tables.}
\end{figure}

Leveraging the geometric configuration illustrated in \cref{fig:InitialSystem} and considering the initial outward velocities of the Ta shells (detailed in Appendix D), we employed the baseline Pb SESAME Table 3206 to generate a representative thermodynamic trajectory. This trajectory pertains to a Lagrangian tracer situated within the Pb material, as presented in \cref{fig:Thermodynamic_Traj}.

As may be observed from examination of \cref{fig:Thermodynamic_Traj}, the hydrodynamic trajectory of the tracer passes through four EoS phases during the course of the simulation. Representative density fields for the simulation using the baseline EoS Sesame Table 3206 for Pb are presented in \cref{fig:3206_DensityFields}.   
\iffalse
\begin{figure}[H]
    \centering
    \includegraphics[width=0.9\textwidth]{Paper_Images_Final/Pb_3206_dynmaics.pdf}
    \caption{Density fields for Pb Sesame Table 3206 g/cm$^3$.} 
    \label{fig:3206_DensityFields}
\end{figure}
\fi
\begin{figure}[H]
    \centering
    \includegraphics[width=0.9\textwidth]{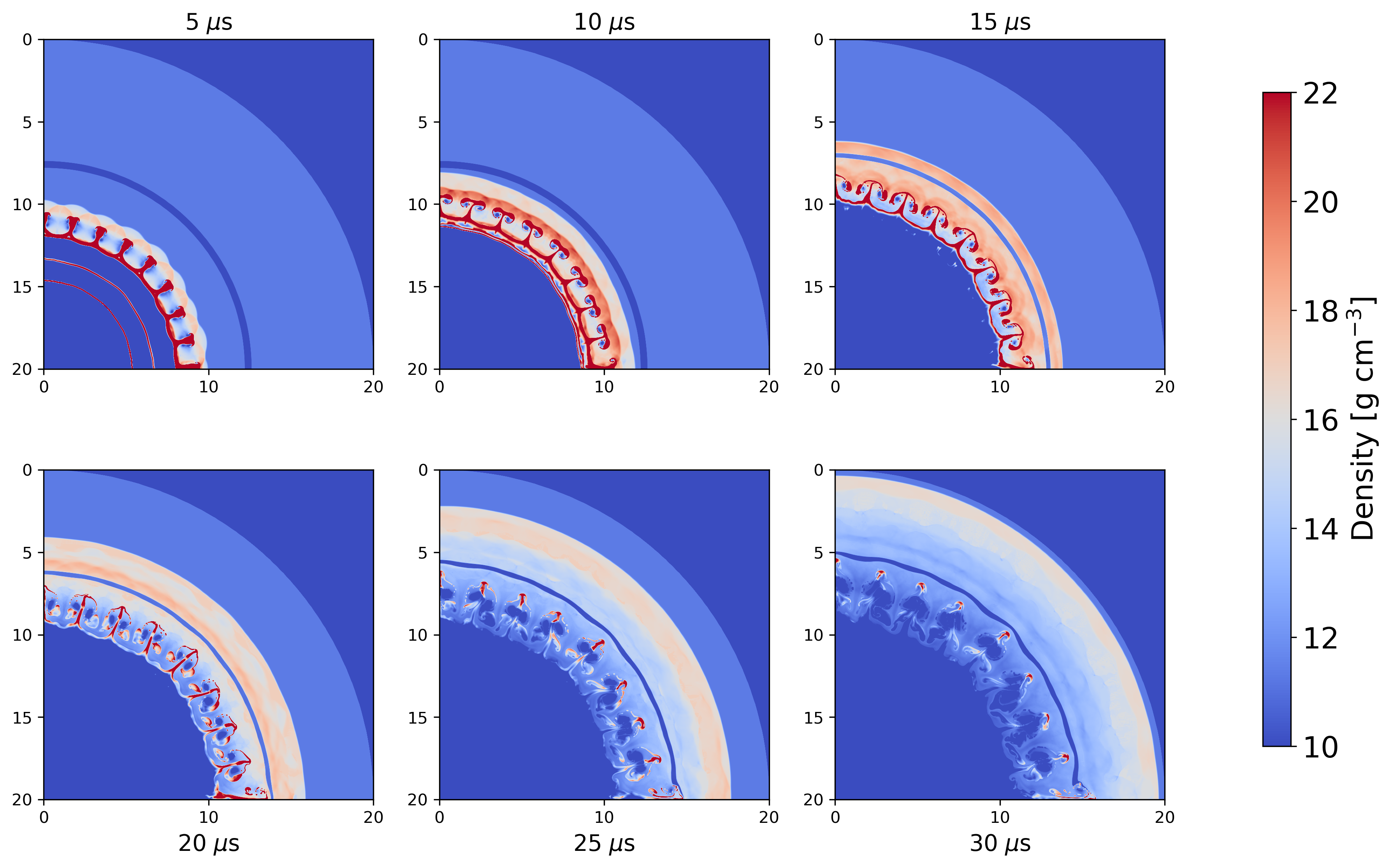}
    \caption{Density fields for Pb Sesame Table 3206 g/cm$^3$.} 
    \label{fig:3206_DensityFields}
\end{figure}

Examination of \cref{fig:3206_DensityFields} indicates the development of a RMI as the shocks generated via the initial outward moving shock interact with the irregular interface at the Ta/Pb interface. Additional complexity in the instability development is achieved by the subsequent interaction of the reflected shocks at the Pb-Cu interface.

\begin{figure}[h!]
    \centering
    \includegraphics[width=0.9\textwidth, trim={0 1.3cm 0 1.3cm}]{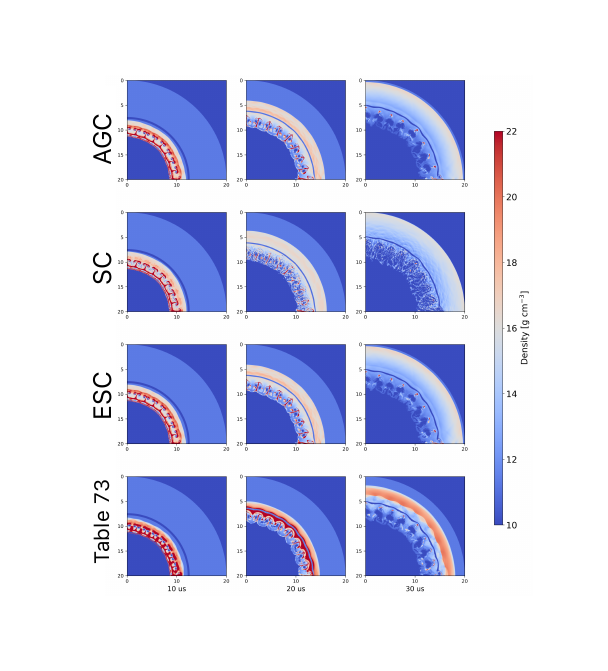}
    \caption{Reconstructed density fields for AGC (first row) and SC (second row) and ESC (third row) Networks along with density field obtained with starting perturbed Pb EoS Table 73 (fourth row}
    \label{fig:3Reconstructed_Table}
\end{figure}

To examine the impact of the reconstructed EoS on the hydrodynamic simulation, we consider a similar setting to that of  \cref{ex:graph_corr}. The ground truth table for Pb (Sesame 3206) is used to compute a set of trajectories for a set of markers $\mfM=\qty{m_i}_{i=1}^M$ in thermodynamic space:
\begin{align}
\mcT=\qty{\mcT_t^{m_i}:\mcT_t^{m_i}=(T_t^i,V_t^i,S_t^i,P_t^i)\in\Phi, m_i\in\mfM, t\in[T_\text{max}]}
\end{align}
for discrete time steps indexed by $t$ that have length $T_\text{max}$. We then map the perturbed Pb Sesame Table (see Appendix C for details) to the ground truth table, using both the AGC and SC corrections, and compare the corresponding hydrodynamic trajectories and density fields to those corresponding to the ground truth table. \cref{fig:3Reconstructed_Table} presents density fields using the AGC, SC, and the  early, mid, and late times.

\begin{figure}[h!]
    \centering
    \includegraphics[width=\textwidth, trim={0 1.3cm 0 1.3cm}]{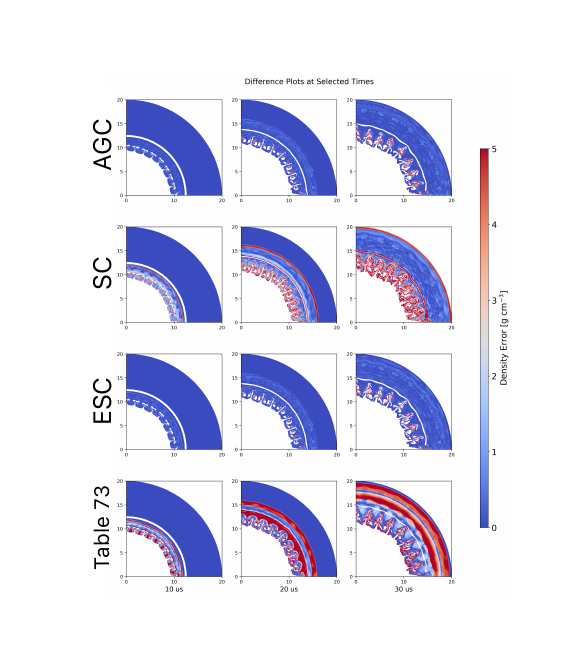}
    \caption{Residuals between the Reconstructed Pb Tables and the Perturbed Pb table with the baseline Pb EoS Table 3206} 
    \label{fig:difference}
\end{figure}

We note that, to perform these simulations we require a completed SESAME table, for which it is necessary to have an estimate of the free energy in addition to $(T,V,P,S)$. This is readily available for AGC but not for the symplectic correction. To remedy this issue, we extend the symplectic network in \cref{app:contact_extension} to compute the induced effect of the SC on the free energy variable. This transformation resembles a Legendre map, and we refer to it as the extended-symplectic correction (ESC). Using this formulation, the free energy can be either completed after fitting the original SC, or fitted along with the other target variables during training.

Several observations are apparent in examining \cref{fig:3Reconstructed_Table}.

1. There are significant differences between the hydrodynamic simulations using the initial Pb EoS perturbed table and those obtained using the Sesame 3206 table

2. Both the AGC and SC networks are able to produce excellent reconstructed Sesame Tables.  Furthermore, when these reconstructed tables are utilized within the frame work of a hydrodynamic simulation excellent matches to both the EoS tables as well as density fields are achieved.

\begin{figure}
    \centering
    \includegraphics[width=0.7\textwidth]{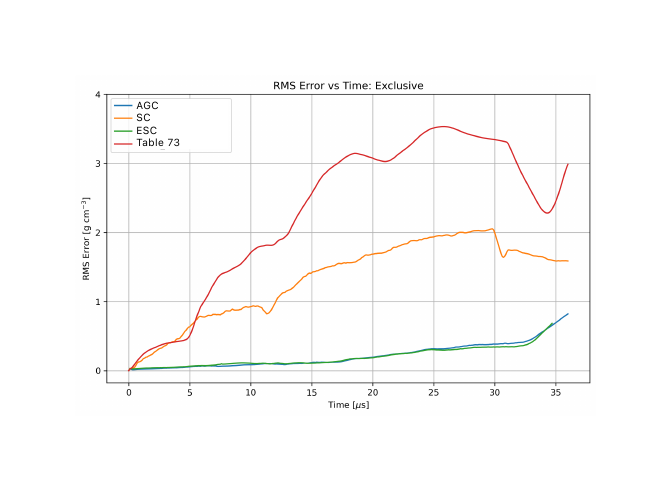}
    \caption{Root-Mean Squared differences between reconstructed and perturbed Eos Tables with baseline Pb EoS Table 3206 } 
    \label{fig:rmserrors}
\end{figure}

Quantitative comparisons of the density differences between the respective tables and SESAME Table 3206, along with those using the perturbed Sesame Table, are provided in\cref{fig:rmserrors} and \cref{fig:difference}.

Finally, comparisons of the respective hydrodynamic paths using the Lagrangian tracer embedded within the Lead are provided in \cref{fig:Thermodynamic_Traj}.  As indicated in \cref{fig:ReconstructPaths}, the tracers using the respective networks, closely follow the path of the baseline table. However, the tracers using the initial perturbed EoS table indicate a somewhat different trajectory within the hcp phase.

\section{Discussion}\label{sec:discussion}

Throughout this work, we propose two computational approaches to representing EoS with neural architectures. Both are capable of incorporating phase transitions in their models, significantly enhancing the applicability of such neural models to applications that may require multiphase EoS. In the case of symplectic networks, this is achieved through \textit{templating}.

Combining the computational capabilities of ML algorithms, in particular neural networks, with EoS allows us to build black- or grey- boxes that fit experimental observations or consistently (in the sense of the first law of thermodynamics) interpolate between observations and theoretical limits. This may be a particular advantage for materials and phase-space regions where performing experiments may be expensive or time-consuming, thus making it difficult to obtain a dense sample of the true EoS.

Our problem approach, that of a `manifold correction', is a particular choice that easily generalizes to other settings. As future work, we propose extensions in (a) the direction of generative EoS modeling (where an appropriate physical model is not available), and (b) the issue of designing proper, general templates that can be used for multiple compounds or elements. Finally, the symplectic correction, which is more flexible in the sense that it can `modify' phase transition regions, is limited since it cannot be applied in a stable manner for multiscale EoS data sets. This is also an important issue to be addressed. Using manifold-type losses to optimize allows movement of phase transition regions.  In this work we do this in an unsupervised manner, in which we allow the network to appropriately move the discontinuity set to minimize the corresponding objective. A supervised approach in the future may be more robust computationally, in which phase boundaries between the input and target EoS are identified prior to optimization and subsequently matched. Furthermore, the manifold-type losses considered in this work are computationally expensive, since distances are computed between large sets of points. Considering approximations of Wasserstein distances between EoS manifolds (using Sinkhorn-type algorithms \cite{cuturi2013sinkhorn}) may allow for faster and more robust optimization of the symplectic network architecture.

\newpage
\bibliographystyle{ieeetr}
\bibliography{bibliography.bib}

\newpage
\appendix
\section{Differential Geometry}\label{app:Differential_Geometry}
\subsection{Theory}\label{app:diff_geo_theory}
\begin{defn}[Symplectic Manifold]
	A manifold $\mcM$ is called symplectic if it has even dimension, and is equipped with a closed non-degenerate differential two-form $\omega$. We denote this as $(\mcM,\omega)$.
\end{defn}

\begin{exmp}
	The usual Euclidean space $\R^{2n}$ parametrized by $(x^1,...,x^n,y^1,...,y^n)$ is symplectic when equipped with the two-form
	\begin{align*}
		\omega=\sum_{i=1}^ndx^i\wedge dy^i
	\end{align*}
\end{exmp}

\begin{defn}[Lagrangian Submanifold]
	Given a symplectic manifold $(\mcM,\omega)$ of dimension $2n$,	a submanifold $\La$ is called Lagrangian if it has dimension $n$ and the restriction of the symplectic form on it vanishes ($\omega\vert_\La=0$).
\end{defn}

In this work, thermodynamic phase space $\Phi$ is treated as a symplectic manifold ($\simeq\R^4$) and the thermodynamic data for any given element or compound lie on a Lagrangian submanifold $\La$ of $\Phi$, a consequence of satisfying conservation of energy. We prove this result in the following section.

\begin{defn}[Symplectomorphism] A symplectomorphism, or symplectic diffeomorphism $\phi$ is a map between two symplectic manifolds $(\mcM,\omega)$ and $(\mcM',\omega')$ that `preserves' the symplectic form, in the sense that the pullback satisfies
\begin{equation*}
\phi^*\omega'=\omega
\end{equation*}
Symplectomoprhisms are diffeomorphisms (smoothly invertible) and volume-preserving maps. In our work, we identify $\omega$ with $\omega'$ since we consider automorphisms of the thermodynamic phase space $\Phi$.

\end{defn}

\subsection{The Geometric Structure of EoS Manifolds}\label{ap:Geometric_structure}
A \textbf{Thermodynamic Variable} is a physical quantity associated with the state of an atomic substance or chemical compound. Examples include Temperature $(T)$, Pressure ($P$), Volume ($V$), Entropy ($S$), Internal Energy ($U$), Helmholtz Free Energy ($A$), Enthalpy ($H$), etc. Each variable is real-valued, and possibly non-negative.

For a given element or compound, these quantities are related, and in general specify a two-dimensional submanifold of Euclidean space. Importanty, such thermodynamic submanifolds are generically \textbf{graphs} of a function, i.e., knowing two of these quantities allows one to infer the rest uniquely. An \textbf{Equation of State} (EoS) is a mathematical model that describes the relationship between these variables.

In general, we may split the variables into three types:

\begin{enumerate}
    \item Independent Variables (e.g., two of $\qty{T,V,P,S,\text{etc.}}$).
    \item Thermodynamic potential functions, (e.g., $\qty{U,H,A,\text{etc.}}$), seen as scalar functions of the independent variables, uniquely determined by them for a given element.
    \item Dependent variables (e.g., the remainder of $\qty{T,V,P,S,\text{etc.}}$) which are functions (in particular derivatives) of the energy functions.
\end{enumerate}

We first establish the relationship between EoS and Symplectic Geometry. Henceforth we use $(T,V,P,S,A)$ as our thermodynamic variables of choice, and we will in particular treat $(T,V)$ as the independent variables, $A$ as the corresponding energy functional, and 
\begin{equation}
    \begin{split}
        S\doteq s(T,V)=-\pdv{A(T,V)}{T}\\
        P\doteq p(T,V)=-\pdv{A(T,V)}{V}
    \end{split}
\end{equation}
as the dependent variables. This is done to demonstrate both theory and applications with some consistence; however, one may reformulate the statements with any other suitable choice of variables. Indeed, a choice of thermodynamic potential usually implies a particular choice of independent variables \cite{mattsson2016short}.

\begin{defn}[Thermodynamic Phase Space]
	Thermodynamic Phase Space $\Phi$ is a 4-dimensional symplectic manifold when parametrized by
	\begin{align}
		\Phi = (T,V,S,P)\in\R^4
	\end{align}
    and equipped with the canonical symplectic two-form
	\begin{align}
		\omega = dT\wedge dS+dV\wedge dP
	\end{align}
\end{defn}

In this space, the \textbf{First Law of Thermodynamics} can be written in differential form as
\begin{defn}[Conservation of Energy]
    Given the symplectic manifold $\Phi$, the first law of thermodynamics can be expressed as the differential one-form
    \begin{equation}
        dA=-SdT-PdV
    \end{equation}
\end{defn}

Furthermore, for a given element or compound, its thermodynamic data lie on a two-dimensional \textbf{Lagrangian submanifold} $\La\subset\Phi$ which satisfies conservation of energy, a consequence of satisfying the first law. We use the term `\textbf{energy-consistent}' to describe submanifolds and models that satisfy this property.

Additionally, we observe that:
\begin{align*}
    0=d(dA)&=-\pdv{S}{V}dV\wedge dT+\pdv{P}{T}dV\wedge dT\\
    &=\qty(\pdv{P}{T}-\pdv{S}{V})dV\wedge dT
\end{align*}
which yields the equivalent constraint to $\omega$ vanishing in the form of a differential equation.

\subsection{Phase Transitions}\label{sec:phase_transitions}
This thermodynamic submanifold $\La$ is assumed to be smooth within a single phase of the compound or material (i.e., the graph of a $C^\infty$ function). However, it is not smooth across phase boundaries, locations in phase space where phase transitions occur, where $\La$ represents a material with multiple phases. Classically, according to the \textbf{Ehrenfest classification} \cite{jaeger1998ehrenfest}, phase transitions are classified as
\begin{enumerate}
    \item 1\textsuperscript{st} Order, when the first derivative of the energy function ($A$) is discontinuous.
    \item 2\textsuperscript{nd} Order, when the first derivative of the energy function ($A$) is well-defined, but its second order derivative is discontinuous.
\end{enumerate}

In the latter case, $\La$ is smooth almost everywhere, and is the graph of a Lipschitz function. Phase transitions appear as `kinks' or `ridges' in projections of the embedded submanifold $\La$, and we also refer to them simply as `discontinuities', due to the ill-defined tangent space at the particular points.

Phase transitions are `interesting' but irregular regions of phase space that must be modelled carefully to obtain an accurate and energy-consistent multiphase EoS.

\subsection{Templating}\label{sec:Templating}
Our proposed approach (\cref{sec:prob_statement}) can be seen through the lens of Templating: Starting from a qualitatively correct, exact model, learning a transformation that maps the model to observed data. In the EoS case, the qualitative feature that must be present in the template is primarily the type (and topology) of phase transitions. However, designing templates is not a trivial task, especially if one expects to find global (multiscale, multiphse) EoS models that accommodate several elements. The main advantage of templating, however, is that one may use continuous techniques to manipulate `discontinuous' templates, effectively removing the need to incorporate the discontinuities into new architectures. They further allow us to use the expressivity of neural networks while retaining some control over interpretable features, as in \cref{ex:analytic}. Additionally, such parametric dependence grants a generative capability, in which parameters can be further \textit{fitted or sampled} after initial training (\cite{GUAN2022114217}). For a general reference on `deformable templates', see \cite{younes2010shapes}.

\section{Symplectic Networks}\label{app:symplectic_networks}
\subsection{Formulation}
Symplectic neural networks are parametric families of functions $f_\theta$ that exactly preserve the symplectic structure of their input (a symplectic manifold) under any choice of parameters $\theta$. In particular, if $\mcM$ is a symplectic manifold, then so is $\mcN\doteq f_\theta(\mcM)$.

In this work we use the \textit{H\'{e}nonNet} architecture introduced in \cite{henonnet}. We restate the following definitions for completeness.

\begin{defn}[H\'{e}non Map]\label{def:henon_map} Given a smooth function $V:\R^{n}\to \R$ and a constant vector $\eta\in \R^n$, a H\'{e}non map is defined as
\begin{align}
    H[V,\eta]:\mqty(x\\y)\mapsto\mqty(y+\eta\\\grad{V}(y)-x)
\end{align}
for $x,y\in\R^n$.
\end{defn}

It is easy to check that both this map and finite compositions of the form $H[V_1,\eta_1]\circ H[V_2,\eta_2]\circ...\circ H[V_n,\eta_n]$ are symplectic, for arbitrary choises of $V_i$ and $\eta_i$.

The underlying result that makes maps of this form particularly useful is that for appropriate $V$ and $\eta$, they are dense within the family of symplectic maps on $\R^{2n}$. This is formally captured in \cite{Turaev_2002}.

In practice, $V$ is the only part of the structure that is estimated by a (deep) feed-forward neural network (FNN). Since FNNs are universal approximators of differentiable scalar functions \cite{cybenko1989approximation, hornik1990universal} $f:\R^n\to\R$, the H\'{e}non Net architecture is capable of approximating any symplectic map.

\begin{defn}[H\'{e}non Layer] Given a H\'{e}non map $H[V,\eta]$, the corresponding H\'{e}non Layer $L[V,\eta]$ consists of the four-fold composition of the original map:
\begin{align}
    L[V,\eta]=H^4[V,\eta]
\end{align}
\end{defn}

Note that this is not a necessary step to design a network-based symplectic map approximator. Nevertheless, the choice of using the layer as a basis block is stable in practice. Observe that for $V\equiv 0$, the $L[0,\eta]$ is the identity map in $\R^{2n}$.

\begin{defn}[H\'{e}non Network] A H\'{e}non network $H_\text{nn}$ of $K$ layers is a map formed by composing $K$ distinct H\'{e}non layers $\vb{L}=\qty{L[V_i,\eta_i]}_{i=1}^K$:
\begin{align}
    H_\text{nn}[\vb{L}]=L_K\circ...\circ L_1
\end{align}
where $L_i=L[V_i,\eta_i]$.
\end{defn}

We also observe that H\'{e}non maps have a closed-form inverse \cite[Remark 2.2]{duruisseaux2022approximation}:

\begin{defn}[H\'{e}non Map Inverse] Given a H\'{e}non map $H[V,\eta]$ it's inverse is the map defined as 
\begin{align}
    H^{-1}[V,\eta]:\mqty(x\\y)\mapsto\mqty(\grad{V}(x-\eta)-y\\x-\eta)
\end{align}

Thus, composing these such inverses in an appropriate manner gives a closed form inverse for any particular instance of a H\'{e}non network architecture. This is convenient computationally since an inverse network is readily available when needed, but also allows us to initialize the network architecture to be the identity map before training (by composing with its original inverse which is not updated). This is useful when the the input manifold is already a good approximation of the target manifold (such as in the EoS case discussed where $\La$ is originally estimated as an approximation to $\La'$).

\end{defn}

\subsection{Contact Extension}
\label{app:contact_extension}
After training the symplectic networks, it may be necessary to lift back to an integral variable (i.e., the energy potential in the case of EoS). This computation is needed, for example, when completing a SESAME table to use for the subsequent hydrocode computations of section \cref{sec:hydrocode}. It is easy to check that the extended transformation
\begin{align}
    L[V,\eta]:\mqty(x\\y\\z)\mapsto\mqty(y+\eta\\\grad V(y)-x\\ z-xy+V(y))
\end{align}
(where $z$ represents the integral variable, i.e. the free energy) is invertible and a contact transformation, restricting to the symplectic transformation when projected onto the $(x,y)$ variables. This can be used either as a post-processing step, or during training (where the generalization to a contact neural network is straight-forward) if integral variable data is available.

\subsection{Manifold Losses}\label{app:manifold_losses}
Aside from the classical pointwise distances between manifolds ($\ell^2$ in the case of descrete samples), we also consider `Manifold losses', which are Hausdorff-like distances between point clouds (that are permutation invariant). Any manifold $\mathcal{M}$ is only sampled by a finite collection of points, and we define our notions of distance accordingly, though there exist consistent formulations in the continuous setting.

Let $\qty{\vb{p}_i}_{i=1}^N=\qty{(\vb{x}_i^p,\vb{y}_i^p)}_{i=1}^N$, with $\vb{x}_i,\vb{y}_i\in \R^{n}$, be a sample of a Lagrangian submanifold $\La\subset\R^{2n}$ (i.e., a discrete collection of points) and define $\qty{\vb{q}_i}_{i=1}^N$ analogously for $\La'$. It is useful to first define:

\begin{align}
    d(\vb q,\mathcal{M})&=\min_{\vb{p}\in\mathcal{M}}\norm{\vb{q}-\vb{p}}_2^2\label{eqn:point_manifold_distance}\\
    d(\mathcal{M},\mathcal{M'})&=\max_{\vb{q}\in\mathcal{M'}} d(\vb p,\mathcal{M})\label{eqn:manifold_manifold_distance}
\end{align}
\cref{eqn:point_manifold_distance} is the classical point-manifold distance, while \cref{eqn:manifold_manifold_distance} is a way of generalizing it to a distance between manifolds. Importantly, it is \textit{asymmetric}. From an optimization perspective, we would like to work with continuously relaxed versions of the $\min$ and $\max$ functions that appear above. To that end we will use the `smooth minimum' function $s_a$ which is continuous and tends to $\min$ in the limit as $a\to\infty$:
\begin{align}
    s_a\qty(\qty{\vb{p}_{i=1}^N})=\frac{\sum_{i=1}^N\vb{p}_ie^{-a\vb{p}_i}}{\sum_{i=1}^Ne^{-a\vb{p}_i}}
\end{align}
We will directly replace the $\max$ with a sum (alternatively an average) to define an equivalent asymmetric manifold distance
\begin{align}
    d_{s_a}(\mathcal{M},\mathcal{M'})=\sum_{i=1}^N s_a\qty(\qty{\norm{\vb{p}_i-\vb{q}_j}_2}_{j=1}^N)
\end{align}

This allows us to define the following general loss functions that can be used as optimization criteria for the symplectic correction of \cref{sec:symp_approach}, whose attributes we discuss below:

\begin{align}
    L_2: &\qquad L_2(\mathcal{M},\mathcal{M}')=\frac{1}{N}\sum_{i=1}^N\norm{\vb{p}_i-\vb{q}_i}^2_2\label{eqn:L2}\\
    L_H: &\qquad L_H(\mathcal{M},\mathcal{M}')=\max\qty{d(\mathcal{M,M'}),d(\mathcal{M',M})}\label{eqn:Hausdorff}\\
    L_H^*:&\qquad L_H^*(\mathcal{M},\mathcal{M}') = d_{s_a}(\mathcal{M,M'})\label{eqn:LH_star}\\
    L_{Hx}^*:&\qquad L_{Hx}^*(\mathcal{M},\mathcal{M}')= d_{s_a}(\mathcal{M,M'}) + \frac{1}{N}\sum_{i=1}^N\norm{\vb{x}_i^p-\vb{\hat{x}}_i^q}^2_2\label{eqn:LHx_star}
\end{align}

The $L_2$ loss (\cref{eqn:L2}) maps points with the same index $i$ between $\La$ and $\La'$. When minimized, it yields a \textit{good} correspondence between the two manifolds pointwise, but comes with two nontrivial drawbacks: (a) It is not invariant to permutations in one sample, so one must be certain of the point-to-point correspondence between the manifold samples before optimizing. This is unfortunate since any permutation of the manifold sample \textit{ought to} define the same geometric object. (b) It does not allow movement of a potential discontinuity (kink), since the index $i$ where a discontinuity would be mapped to is fixed, predefined by the point-to-point correspondence of manifold samples. The first hurdle is overcome by the more natural Hausdorff-like distance $L_H$ (\cref{eqn:Hausdorff}), which is invariant to permutations (at the cost of being more expensive to compute). It does not directly solve the discontinuity-movement issue.

To address both we consider the asymmetric (denoted by $^*$) Hausdorff-like distances $L_H^*$ (\cref{eqn:LH_star}) and $L_{Hx}^*$ (\cref{eqn:LHx_star}). Aside from the fact that they are differentiable through the modification of the $\min$ and $\max$ functions of $L_H$, the removal of the symmetry constraint in $L_{H}^*$ allows discontinuities to move more freely in ambient space. That comes at the cost of having small residuals even if points in $\La$ are far from points in $\La'$ (but not conversely). To counteract this effect, we introduce a Lagrangian stabilization term in $L_{Hx}^*$ which, under a symplectomorphism $f:\La\mapsto\La'$ penalizes large changes in the Lagrangian subspace (where $\hat{x}_i^q$ is the $x$-coordinate of the image $f(\La)$). This choice comes from the knowledge that $\vb{y}_i^p$ is ultimately a function of $\vb{x}_i^p$, so it is natural to ask for an adjustment of only the latter half of the coordinates, and this is further justified as a modelling choice in the context of \eos (\cref{ap:Geometric_structure}) under the independent-variable assumption. 

We note that the two individual terms of $L_{Hx}^*$ may compete in their objectives: In their relative lexicographic limits, they effectively approximate $L_H^*$ and $L_2$ respectively. However, our empirical results suggest that this competition can be critically useful when the terms are weighted appropriately. The following toy example offers some computational evidence of the preceding discussion.

\begin{exmp}[1D Toy]\label{exmp:toy}
Let $x\in\qty[-2,2]$ denote the independent variable, and consider two piecewise differentiable potential functions $V$ and $U$ given by:
\begin{gather}
    V(x)=\frac{\sgn(x)x^2}{2}+x\\
    U(x) =\frac{\sgn(x-1)(x-1)^2}{2}+1
\end{gather}

Then, define the piecewise smooth Lagrangian submanifolds $\La'=\qty{(x,v)}\equiv\qty{(x,y):y=V_x}$ and $\La=\qty{(x,u)}\equiv\qty{(x,y):y=U_x}$ with the canonical two-form $\omega=\dd{x}\wedge\dd{y}$, which is well-defined everywhere except from a set of measure zero on each (\cref{fig:2d_discontinuity}). 

\begin{figure}
\centering
\includegraphics[width=0.5\textwidth]{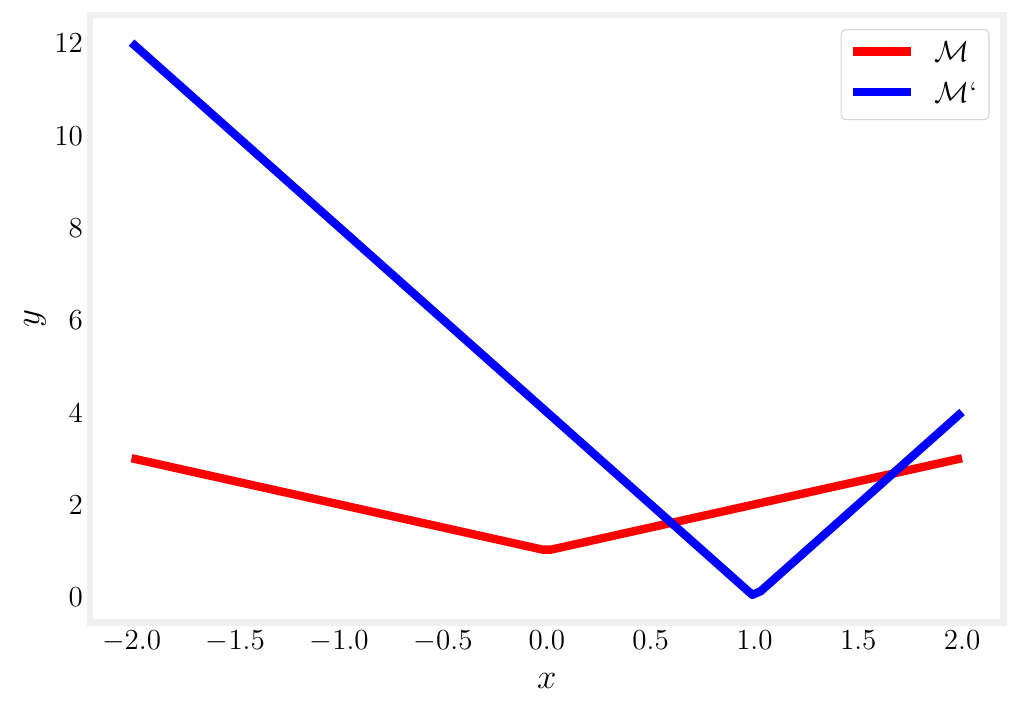}
\caption{Graphical depiction of the piecewise smooth Lagrangian submanifolds $\La'$ (\textcolor{red}{red}) and $\La$ (\textcolor{blue}{blue}).}
\label{fig:2d_discontinuity}
\end{figure} 

We say that $\La$ approximates $\La'$ in the sense that they have the same topology; however the location of their discontinuity in the derivative (kink) and the \textit{measure}\footnote{whether discrete or Lebesgue, induced by a uniform measure on the independent variable $x\in[0,2]$} of each manifold on either side of it are different. To learn the correction between $\La$ and $\La'$, we sample each manifold using an equally spaced grid of 100 points in $x$, and train a symplectic network $\fnn$ under the various loss functions listed in \cref{app:manifold_losses}. This simple example demonstrates the issues discussed there more abstractly.

In \cref{fig:2d_discontinuity_training} we show snapshots of the training process of a symplectic network $\fnn$ attempting to match $\La$ to $\La'$ under the $L_{Hx}^*$ loss, at different training epochs $t$. At each $t$ we observe that $\La'$ and $\fnn(\La)$ have the same topology, i.e., both are smooth aside from one point where their derivative does not exist, denoted by $\star$. Remarkably, we see that under this loss the resulting network maps the original discontinuity of $\La$ close to the one of $\La'$ in a completely unsupervised manner (with an approximate error $\ell_2$ error of 0.1!). This also implies the concentration of the sample points to the left of the discontinuity (due to the measure imbalance), and a spread of the ones to the right (due to the Lagrangian term of the loss).

\begin{figure}[ht]
    \centering
    \includegraphics[width=0.9\textwidth]{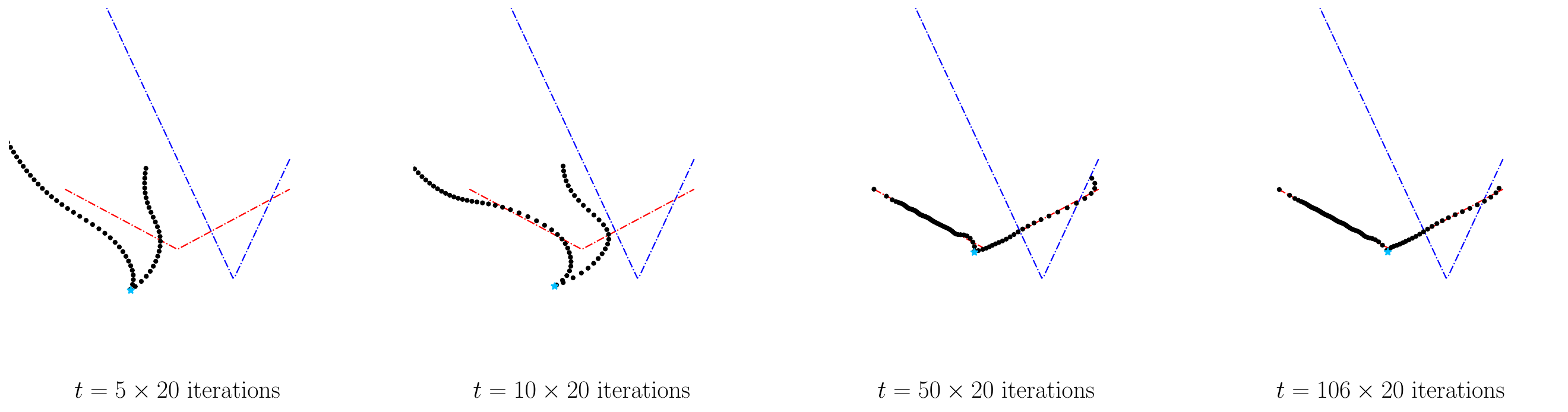}
    \caption{Snapshots of training process under $L^*_{Hx}$. The input to the symplectic network is the manifold $\La$ (\textcolor{blue}{blue}), while the output (black) is matched to the target manifold $\La'$ (\textcolor{red}{red}). The star ($\color{cyan}\star$) denotes the image of the point where the discontinuity in the derivative occurs.}
    \label{fig:2d_discontinuity_training}
\end{figure}

\cref{fig:2d_different_losses} depicts `good' converged results after training the same symplectic network with different loss functions ($L_2,L_H^*,L_{Hx}^*$). We see that $L_2$ does not move the discontinuity at all while covering the full target manifold $\La'$. Contrastingly, $L_H^*$ correctly matches the discontinuities while leaving part of the target manifold (right side of $\La'$ in \cref{ex:2d_LH_star}) uncovered (which may imply extrapolation error). A good balance between the two behaviors is achieved by minimizing $L_{Hx}^*$ (\cref{ex:2d_LHx_star}). On an elementary level, this variety in the density of $\fnn(\La)$ which allows both objectives to be satisfied in the latter case is directly caused by the competing terms in the $L_{Hx}^*$ loss. (It is interesting to compare the result of minimizing of $L_{Hx}^*$ in \cref{fig:2d_different_losses,fig:2d_discontinuity_training}). In \cref{tbl:2d_loss_comparison} we compare the numerical values of all losses after minimizing each individual one (denoted in the right column). The values across each line should be compared vertically and not horizontally, since they make use of different statistics when computing distances. Overall, we observe the numerical evidence of the preceding qualitative discussion: $L_2$ is competitive across all metrics aside from matching the discontinuity points of $\La$ and $\La'$, with a good middle ground in performance achieved by optimizing $L_{Hx}^*$.

At this point, we emphasise the use of the word `good' in the previous paragraph. The optimization results (of e.g., \cref{fig:2d_different_losses}) are sensitive to both the initialization of the network and the optimization procedure in the following manner: The asymmetric manifold distance is minimized when $\fnn(\La)$ is in the interior of $\La'$, so in principle the black points can be folded to only fit on a small, non-representative part of $\La'$. Increasing the weight of the Lagrangian term counteracts this issue, but comes at a cost of being less sensitive to topology. Additionally, large learning rates are capable of distorting ambient space very quickly and are hard to invert once a local minimum is achieved. This phenomenon can be reduced by using smaller learning rates. Importantly, the sole reason we are able to match the discontinuities of the two manifolds is that it is `expensive' for our symplectic networks to create or destroy them. However, it is capable of doing so (as seen e.g., in \cref{ex:2d_L2}) and if that occurs, topological characteristics can then be lost (in the sense that they are no longer factored in by the loss).

Despite these issues, we note that that it is not unreasonable to mark and match discontinuities in a supervised manner, and to have a good approximation of $\mathcal{M}$ as an initial guess. That information, depending on the application, may often be tractably available, and considerably reduces the computational issues discussed.

\begin{figure}[!h]
  \begin{subfigure}[b]{0.33\textwidth}
    \includegraphics[width=\textwidth]{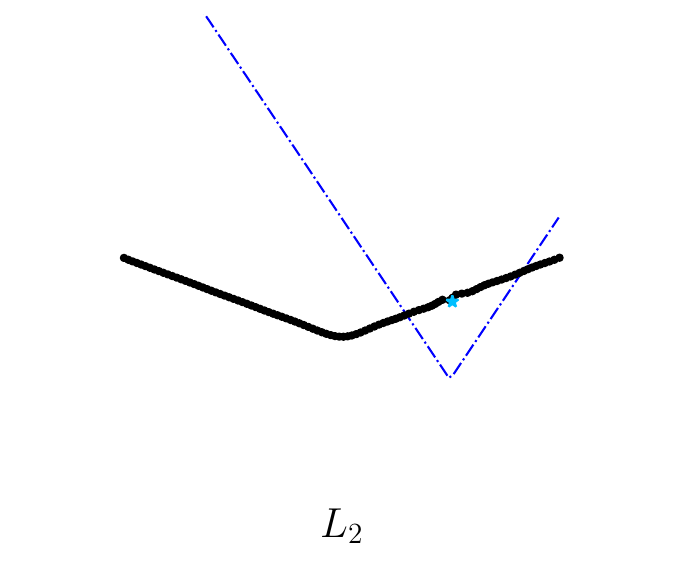}
    \caption{}
    \label{ex:2d_L2}
  \end{subfigure}
  \begin{subfigure}[b]{0.33\textwidth}
    \includegraphics[width=\textwidth]{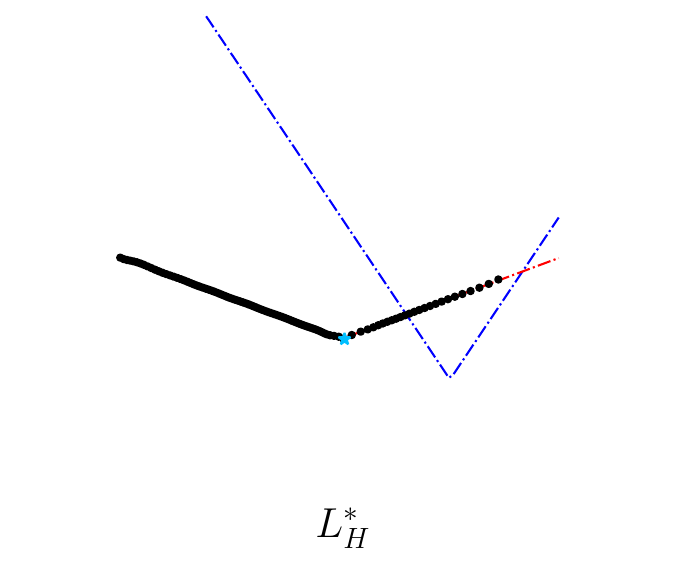}
    \caption{}
    \label{ex:2d_LH_star}
  \end{subfigure}
  \begin{subfigure}[b]{0.33\textwidth}
    \includegraphics[width=\textwidth]{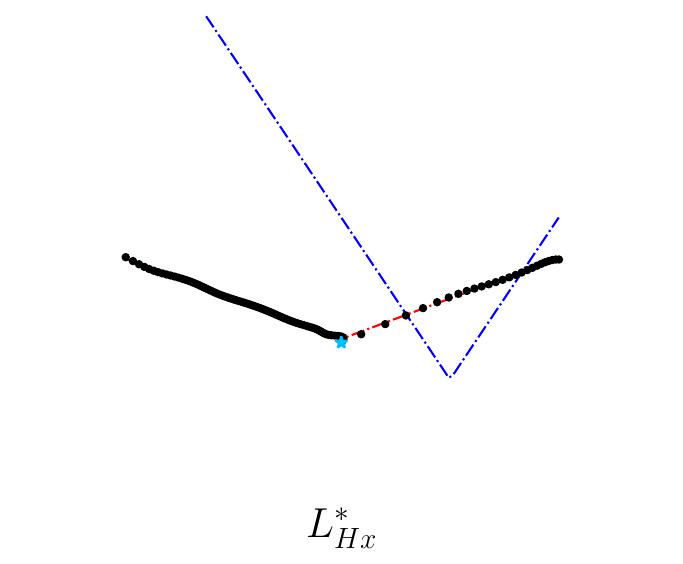}
    \caption{}
    \label{ex:2d_LHx_star}
  \end{subfigure}
  \caption{Mappings of $\mathcal{M}'$ (\textcolor{blue}{blue}) to $\mathcal{M}$ (\textcolor{red}{red}) obtained using $\fnn$ with different choices of loss function. The image of the point where the discontinuity occurs in $\mathcal{M'}$ is denoted by a star ($\color{cyan}\star$).}
    \label{fig:2d_different_losses}
\end{figure}

\begin{table}
\centering
\begin{tabular}{cccccc}
% \toprule
Objective & $L_2$ & $L_H$ & $L_H^*$ & $L_{Hx}^*$ & Discontinuity \\
\midrule
$L_2$      & \cellcolor{white!10}\textbf{2 e-4}& 0.048 & 0.101 & 0.101 & \cellcolor{white!20}\xmark\\
$L_H^*$    &  0.585 &0.132&\cellcolor{white!10}\textbf{0.106} &0.815&\cellcolor{white!20}\checkmark\\
$L_{Hx}^*$ &  0.373  & 0.077& 0.101 &\cellcolor{white!10}\textbf{0.568} & \cellcolor{white!20}\checkmark\\
\bottomrule
\end{tabular}
\caption{Loss comparison corresponding to \cref{fig:2d_different_losses}. The optimized objective (left column) compared to the other loss values evaluated on the \textit{test} set after convergence of the neural network. The right-most column qualitatively states if the discontinuity points of are matched.}
\label{tbl:2d_loss_comparison}
\end{table}

\end{exmp}

\subsection{Establishing Lipschitz Control}
The previous example (\cref{exmp:toy}) demonstrates that coupling H\'{e}non networks with a manifold loss can implicitly `identify' and move discontinuities. Since the networks are diffeomorphisms, they are guaranteed to preserver the local differential structure of submanifolds, and will preserve a discontinuity of their input; however, they are capable of smoothing out such a discontinuity, and alternatively warping a smooth part of the input to approximate a discontinuity. This behavior is visible in \cref{fig:2d_different_losses}, where, under the $L_2$ loss, the network is `forced' to smooth out the area surrounding the kink.

In this section we demonstrate that H\'{e}non networks are Lipschitz maps, and propose control of their Lipschitz constant to improve training stability. The Lipschitz constant is an additional meta parameter of the network and must be chosen appropriately for the problem.

\subsubsection*{Formal Statement}
\begin{prop}
A H\'{e}non map is $(l, L)$-bi-Lipschitz where $l, L$ are constants depending on $\grad V$. That is, there are constants $l,L$ that satisfy
\begin{align*}
\frac{1}{l} d_X\qty(\mqty(x'\\y'),\mqty(x\\y))
\leq
    d_Y\qty(H[V,\eta]\mqty(x'\\y'),H[V,\eta]\mqty(x\\y))
    \leq L d_X\qty(\mqty(x'\\y'),\mqty(x\\y)),
\end{align*}
for arbitrary metrics $d_Y,d_X$ on $\R^n$, for which we will generally use $\norm{\cdot}$ in place of the Euclidean norm $\norm{\cdot}_2$. 

\proof
Given $x,y,\eta\in\R^n$ and a (smooth) potential function $V:\R^n\to\R$, we defined
\begin{align*}
    H[V,\eta]:\mqty(x\\y)\mapsto\mqty(y+\eta\\-x+\grad{V}(y))
\end{align*}
in \cref{app:symplectic_networks}. 
We have that:
\begin{align*}
    H[V,\eta]\mqty(x\\y)-H[V,\eta]\mqty(x'\\y')&=\mqty(y'-y\\x-x'+\grad{V}(y')-\grad{V}(y))\\
    &=\mqty(0&I\\-I&G\qty(y,y'))\mqty(x'-x\\y'-y)
\end{align*}
where $G(y,y')\in\R^{n\times n}$ is a matrix function that satisfies:
\begin{align}
    \grad{V}(y')-\grad{V}(y)=G(y,y')(y'-y)
\end{align}
Thus, $L$ is given by
\begin{align*}
    L\doteq \sup_{y,y'\in\R^n}\norm{\mqty(0&I\\-I&G(y, y'))}
\end{align*}
We can bound $L$ in terms of the norm of $G$ as follows: For arbitrary matrices $\mcM\in\R^{2n\times 2n}\qc\mcG\in\R^{n\times n}$ such that
\begin{align}
    \mcM=\mqty(0&I\\-I&\mcG)
\end{align}
the euclidean norm of $\mcM$ is bounded:
\begin{align*}
    \norm{\mcM}^2=\sup_{\norm{x}^2+\norm{y}^2=1}\norm{\mcM\smqty(x\\y)}^2&=\sup_{\norm{x}^2+\norm{y}^2=1}\norm{\mqty(y\\\mcG y-x)}^2\\
    &=\sup_{\norm{x}^2+\norm{y}^2=1}\qty{\norm{y}^2+\norm{\mcG y-x}^2}\\
    &\leq\sup_{\norm{x}^2+\norm{y}^2=1}\qty{\norm{x}^2+\norm{y}^2+\norm{\mcG}^2\norm{y}^2}\\
    &\leq 1+\norm{\mcG}^2
\end{align*}
The above bound is tight since it is achieved when $x=0,y=\arg\sup_{\norm{y}=1}\norm{\mcG y}$, where the operator norm is defined as  $\norm{\mcG}=\sup_{\norm{z}=1}\norm{\mcG z}\qq{for} z\in\R^n$. Thus, the Lipschitz constant for a H\'{e}non transformation is given by 
\begin{align}
    L=1+\sup_{y,y'}\norm{G(y,y')},
\end{align}
provided that $\norm{G(y,y')}$ is bounded.

In a similar fashion, we can find the Lipschitz constant, $l$, of the
inverse map,
\begin{align*}
    H^{-1}[V,\eta]:\mqty(x\\y)\mapsto\mqty(\grad{V}(x-\eta)-y\\x-\eta),
\end{align*}
which can be shown to be $l = L$.

\newcommand{\remove}[1]{}
\remove{
and we are interested in computing this quantity for various choices of $V$, which in our implementation is a feed-forward neural network.

We note that a similar calculation can be done for the inverse of a H\'{e}non transformation. 
The inverse map is given by
\begin{align*}
    H^{-1}[V,\eta]:\mqty(x\\y)\mapsto\mqty(\grad{V}(x-\eta)-y\\x-\eta).
\end{align*}
We have
\begin{align*}
    H^{-1}[V,\eta]\mqty(x\\y)-H^{-1}[V,\eta]\mqty(x'\\y')&=\mqty(\grad{V}(x-\eta)-\grad{V}(x'-\eta)-y+y'\\x-x')\\
    &=\mqty(F(x,x')&I\\-I&0)\mqty(x'-x\\y'-y)
\end{align*}
where $x,y,x',y',\in\R^n$ and $F(x,x')$ satisfies
\begin{align*}
    \grad{V}(x-\eta)-\grad{V}(x'-\eta)=F(x,x')(x'-x)
\end{align*}
Note that $F(x, x') = -G(x-\eta, x'-\eta)$.
Thus
\begin{align}
    l \doteq\sup_{x,x'\in\R^n}\norm{\mqty(F(x, x')&I\\-I&0)}
\end{align}
and for an arbitrary matrix $\mathcal{F}\in\R^{n\times n}$,
\begin{align*}
    \sup_{\norm{x}^2+\norm{y}^2=1}\norm{\mqty(\mathcal{F}x+y\\-x)}^2&=\sup_{\norm{x}^2+\norm{y}^2=1}\qty{\norm{\mathcal{F}x+y}^2+\norm{x}^2}\\
    &\leq\sup_{\norm{x}^2+\norm{y}^2=1}\qty{\norm{x}^2+\norm{y}^2+\norm{\mathcal{F}}^2\norm{x}^2}\\
    &\leq 1+\norm{\mathcal{F}}^2
\end{align*}
Once again, the bound is tight since equality is achieved when $y=0\qq{and}x=\arg\sup_{\norm{x}=1}\norm{Fx}$.
}

Overall, these results imply that as long as $\norm{G(y,y')}$,
which depends on $\grad{V}$, is bounded, The H\'{e}non transformation is \textit{bi-Lipschitz} with constants $L,l=L$.
\end{prop}

An immediate consequence of the preceding proposition is that H\'{e}non layers and networks, which consist of compositions of H\'{e}non maps, are also bi-Lipschitz

\begin{prop}
    H\'{e}non layers and networks are bi-Lipschitz maps, with their constants dependent on $\grad{V}$, $\qty{\grad{V_i}}_{i=1}^K$ respectively.
\end{prop}

\subsubsection*{Numerical Implementation}

The bi-Lipschitz H\'enon map
is implemented using a similar approach to the
bi-Lipschitz affine transformation (bLAT) layer 
introduced in~\cite{serino2024}.
The function, $V:\mathbb{R}^{n}\rightarrow 1$ is represented as an arbitrary width 
neural network,
\begin{align}
    V(y) &= c^T \theta(\mathcal{U}\Sigma \mathcal{V}^T y +  b),
\end{align}
where 
$\mathcal{U}\in \mathbb{R}^{k \times r}, 
\Sigma\in \mathbb{R}^{r \times r}, 
\mathcal{V}\in \mathbb{R}^{n \times r}, 
b\in \mathbb{R}^k, 
c\in\mathbb{R}^k$,
$r=\min(k, n)$,
and $\theta$ is an element-wise activation function.
We construct 
$\mathcal{U},\Sigma, \mathcal{V}$
so that 
$A = \mathcal{U}\Sigma \mathcal{V}^T$ is a singular value
decomposition.
Therefore, $\mathcal{U}, \mathcal{V}$ have orthogonal columns
and $\Sigma$ is a diagonal matrix of singular values,
$\Sigma={\rm diag}(\sigma_1, \dots, \sigma_r)$.
Furthermore, we restrict the singular values
$|\sigma_i|\le \sigma_{\max}$ for some parameter $\sigma_{\max}>0$.
We additionally construct $c$ such that the maximum element
$\max_j |c_j| \le c_{\max}$ for some parameter $c_{\max} > 0$.

Using index notation, the function $V$ can be represented as
\begin{align}
   V(y) &= \sum_{i} c_i \theta\left(\sum_{j} A_{ij} y_j + b_i\right).
\end{align}
The $k^{\rm th}$ component of the gradient is given by
\begin{align}
    \frac{\partial}{\partial y_k} V(y) = 
    \sum_{i} c_i A_{ik} \theta'\left(\sum_{j} A_{ij} y_j + b_i\right).
\end{align}
Let $z_i = \sum_{j} A_{ij} y_j + b_i$ and $z_i' = \sum_{j} A_{ij} y_j' + b_i$ Then
\begin{align}
    \frac{\partial}{\partial y_k} V(y') - \frac{\partial}{\partial y_k} V(y) &= 
    \sum_{i} c_i A_{ik} \left(\theta'\left(z_i'\right) - \theta'\left(z_i\right)\right), \nonumber \\
    &= \sum_{i} c_i A_{ik} \lim_{\epsilon\rightarrow 0}\frac{\theta'\left(z_i'\right) - \theta'\left(z_i\right)}{z_i' - z_i + \epsilon} (z_i' - z_i), \nonumber\\
    &= \sum_{i} c_i A_{ik} \lim_{\epsilon\rightarrow 0}\frac{\theta'\left(z_i'\right) - \theta'\left(z_i\right)}{z_i' - z_i + \epsilon} \sum_{j} A_{ij} (y_{j}' - y_j).
\end{align}
Therefore the $(k,j)$ element of $G(y', y)$ is given by
\begin{align}
    G(y', y) \big{|}_{kj} = \sum_{i} c_i A_{ik} A_{ij} \lim_{\epsilon\rightarrow 0} \frac{\theta'\left(z_i'\right) - \theta'\left(z_i\right)}{z_i' - z_i + \epsilon}.
\end{align}
This can be written in matrix notation using
\begin{align}
    G(y', y) = A^T {\rm diag}(c) \, {\rm diag}\left(\lim_{\epsilon\rightarrow 0} \frac{\theta'\left(z_i'\right) - \theta'\left(z_i\right)}{z_i' - z_i + \epsilon}\right) A.
\end{align}
Here, ${\rm diag}(c) = \left[\begin{array}{ccc} c_1 &\dots & c_k \end{array}\right]^T$, etc.
Therefore, 
\begin{align}
    \|G(y', y)\|_2 \le \kappa_{\theta'} \|A\|_2^2  \max_j |c_j| ,
\end{align}
where
\begin{align}
    \kappa_{\theta'} := \sup_{z, z'}\lim_{\epsilon\rightarrow 0}\frac{|\theta'(z') - \theta'(z)|}{|z' - z + \epsilon|}.
\end{align}
Due to the mean value theorem,
\begin{align}
    \kappa_{\theta'} \le \sup_{z} |\theta''(z)|.
\end{align}
This constant can be determined analytically or numerically
for different choices of activation functions.
For $\theta=\tanh$, we have 
$\kappa_{\theta'}=\frac{4}{3\sqrt{3}} \approx 0.7698$.

By construction, we have that $\|A\|_2 = \sigma_{\max}$
and $\max_j |c_j| = c_{\max}$.
Therefore, the Lipschitz constant for the H\'enon layer involving
$V$ satisfies
\begin{align}
    L = 1 + \kappa_{\theta'} \sigma_{\max}^2 c_{\max}.
\end{align}

We note that in our implementation of the layer,
the orthogonal matrices are parameterized
using the Householder factorization.

\section{Architectures and Data Sets}\label{app:data_architectures}

\subsection{SESAME tables}
In our numerical examples, we utilize tabular EoS data generated with OpenSesame. In particular, we use `ground truth' tables for lead (Pb) and Copper (Cu) along with a perturbed table for lead. 

OpenSesame is a sophisticated code system specifically designed for developing and interacting with the Los Alamos National Laboratory's SESAME Equation of State (EoS) libraries. These libraries serve as a comprehensive collection of thermodynamic properties of materials, detailed in tables that elucidate the behavior of materials under a vast range of conditions. Central to OpenSesame is the utilization of the Helmholtz free energy as the foundational thermodynamic variable. This choice enables the creation of EoSs that represent material behavior across extensive temperature and pressure ranges. The system is essential for producing EoSs that are both theoretically robust, integrating advanced models and computational techniques, and practically valuable, closely aligning with experimental data and detailed calculations.

OpenSesame employs a three-term decomposition approach to produce EoSs that are reasonably accurate over broad temperatures and pressures. This method decomposes the total Helmholtz free energy, $F(V,T)$, into three distinct components: First, the cold curve, $\phi_0(V)$, encapsulates the internal energy's relationship with volume, absent thermal contributions. Next, the nuclear model, $F_\text{ion}(V,T)$, extends the cold curve by incorporating thermal ionic contributions, capturing the thermal effects on the atomic nuclei. Finally, the electronic model, $F_\text{el}(V,T)$, addresses the electronic contributions to the free energy. This last component is particularly crucial for metals, plasmas, and certain high-pressure/temperature phases of materials, playing a key role in modeling compressibility, thermal expansion, and phase stability under extreme conditions.

By summing these components at constant density ($\rho$) and temperature ($T$), OpenSesame generates a detailed $\rho$, $T$ grid. This grid forms the basis for evaluating all other thermodynamic quantities, providing a comprehensive and nuanced understanding of material behavior across diverse environments. The decomposition approach ensures that OpenSesame can generate accurate and thermodynamically consistent EoSs, catering to a wide spectrum of scientific and engineering applications.

To evaluate the ability of our machine learning approaches to accurately produce EoS Tables we have selected to examine a system in which Lead (Pb) is present.  These materials incorporate a variety of phases and models to accurately describe their physical properties under different conditions.

Pb is characterized by three solid phases—Body-Centered Cubic (BCC), Face-Centered Cubic (FCC), and Hexagonal Close-Packed (HCP)—and one liquid phase. The solid phases utilize the finite-strain model for the cold curve, indicating the strain behavior under stress. The electronic contributions are calculated using the Thomas-Fermi-Dirac (TFD) model, which incorporates quantum mechanical effects of electrons. The nuclear model, giknuc, bridges Debye's low-temperature harmonic oscillations and an ideal gas's high-temperature behavior. The transition from solid to liquid phases employs the Lindemann melt model, capturing the melting process based on atomic vibrations. For the liquid phase, the models used are finite-strain for structural behavior, TFD for electronic aspects, and HighTLiq—a model designed for high temperature liquid phases in a multiphase approach.

A sensitivity study was conducted to identify the parameters most influential in deviating from the base EoS, focusing on the FCC phase for and Pb, while also determining the range of each individual parameter that would produce a valid EoS. The parameters identified for potential adjustments were the FCC \texttt{cold\_bulk\_modulus}, liquid \texttt{cold\_dbdp}, and FCC \texttt{reference\_gamma}. The \texttt{cold\_bulk\_modulus} measures resistance to compression, derived from the pressure-volume relationship at low temperatures. The \texttt{cold\_dbdp} represents the pressure derivative of the bulk modulus at the reference density, indicating how the modulus changes with pressure. The \texttt{reference\_gamma} is applied in the nuclear model to describe low-temperature behaviors.

With these parameters in mind, an ensemble of EoSs for Pb was generated through random sampling of the three identified parameters within the valid range. This approach allows for a comprehensive study of each parameter's impact on the material’s properties. The initial, baseline parameters for Pb are [\texttt{cold\_bulk\_modulus}, \texttt{cold\_dbdp}, and \texttt{reference\_gamma}]= [46.903, 4.7, 2.6944]. Notably, the 73rd table for Pb was selected for further analysis in this work due to its significant deviations in hydrodynamic results when compared to the baseline EoS, with the values as a percentage of the base parameters $[134.61, 51.08, 143.75]$\%.

The tables list the values of $\qty{T_i,V_j,S_{ij},P_{ij},A_{ij}}$ over a grid specified in the independent variables $\qty{T_i,V_j}$. The range of the independent  variables ($T,V$) considered in simulations is listed in \cref{tab:data_units_ranges}, along with typical values of the dependent variables.

\begin{table}[]
    \centering
    \begin{tabular}{ccc}
         Temperature & $K$ & [131, 36426]\\
         Volume & $m^3/Mg$& [0.016, 0.100] \\
         Entropy & $MJ/kg K$ &[117e-3, 4.04e-5]\\
         Pressure & $GPa$ & [2610, 1.67e-8]\\
         Free Energy & $MJ/kg$ & [11, -32]
    \end{tabular}
    \caption{Observables, units, and value ranges (of the independent variables) used in the numerical simulations involving SESAME tables. The dependent variable ranges vary based on the data set.}
    \label{tab:data_units_ranges}
\end{table}

\subsection{Admissible Regularization}
 Translations of data sets are admissible. However, we may not arbitrarily scale each individual variable since that distorts the symplectic structure of thermodynamic phase space $\Phi$. Observe that for given coordinates of a symplectic manifold $\qty{x^1,x^2,y^1,y^2}$, the following two scaling-group actions leave the symplectic form invariant:
\begin{align*}
    \qty{x^1,x^2,y^1,y^2}\mapsto\qty{\alpha x^1,x^2,\frac{1}{\alpha}y^1,y^2}\qc\alpha\in\R^+\\
    \qty{x^1,x^2,y^1,y^2}\mapsto\qty{x^1,\beta x^2,y^1,\frac{1}{\beta} y^2}\qc\beta\in\R^+
\end{align*}
while the following three change the symplectic form by a constant multiplicative factor:

\begin{align*}
    \qty{x^1,x^2,y^1,y^2}&\mapsto\qty{\gamma x^1,\gamma x^2,y^1,y^2}\qc\gamma\in\R^+\\
    \qty{x^1,x^2,y^1,y^2}&\mapsto\qty{x^1,x^2,\delta y^1,\delta y^2}\qc\delta\in\R^+\\
    \qty{x^1,x^2,y^1,y^2}&\mapsto\qty{\zeta x^1,\zeta x^2,\zeta y^1,\zeta y^2}\qc\zeta\in\R^+
\end{align*}

Thus, any combination of these transformations can be used to regularize thermodynamic data (which are of the form $\qty{T_i,V_i,S_i,P_i}_{i=1}^N$ in our numerical experiments. In our computational examples, we typically use values between ($10^{-3},10^{-4}$) for $\alpha$ (multiplying temperature and entropy) and $(10^1,10^2)$ for $\beta$ (multiplying volume and pressure).

\subsection{Computational Summary}
We summarize each computational example in \cref{tbl:computation Results}, where we note the type of model, objective optimized, and final discrepancy when optimization was stopped. The numerical values depend on the example, regularization, and the approximation parameter ($a$, as specified in \cref{app:manifold_losses}). Notably, the symplectic model in \cref{sec:hydrocode} does not perform as well as the extended model which also accounts for free energy values during training using the contact extension of \cref{app:contact_extension}.

\begin{table}[H]
\centering
\begin{tabular}{lccc}
Example & Architecture & Optimized Loss & Final Loss\\
\midrule
\cref{ex:graph_corr}& \makecell[c]{Graph \\ Additive Graph} & $\ell^2$ &\makecell[c]{3.57e-2\\1.15e-2}\\
\midrule
\cref{ex:symp_corr}& Symplectic &\makecell[c]{$\ell^2$\\$L_{Hx}^*$} &\makecell[c]{0.27e-4\\8.45e-6}\\
\midrule
\cref{ex:Cu_to_Pb}& Symplectic & $L_{Hx}^*$ &9.62e-1\\
\midrule
\cref{ex:analytic} & Symplectic & $L_{Hx}^*$ &8.77e-3\\
\midrule
\cref{sec:hydrocode}& \makecell[c]{Additive Graph\\Symplectic\\Contact} & \makecell[c]{$\ell^2$\\ $L_{Hx}^*$\\$L_{Hx}^*$} &\makecell[c]{2.92e-2\\1.23\\1.11e-2}\\
\bottomrule
\end{tabular}
\caption{}
\label{tbl:computation Results}
\end{table}

\subsection{Architecture Specifications}
The mathematical description of the neural architectures used is described in \cref{sec:methods}. Here we specify the exact parameters by which they are implemented in each numerical example.

All networks are implemented in \texttt{python} using \texttt{pytorch} \cite{torch_NEURIPS2019_9015} and the default weight initialization is used. All (individual) network implementations are fully connected, with a combinaiton of $\tanh$ and linear activation functions, as specified in \cref{tbl:architectures}. The networks are optimized using Adam \cite{kingma2014adam}. All instances of H\'{e}non networks were constructed by composing three H\'{e}non layers.

\begin{table}[H]
\centering
\begin{tabular}{ccc}
% \toprule
Example & Estimated Function & Layers \& Activation, Width \\
\midrule
\cref{ex:graph_corr} & $\hat{f}_G$ & (4 $\tanh$, 1 linear), 20\\
\cref{ex:graph_corr} & $\hat{f}_G^+$ & (4 $\tanh$, 1 linear), 20 \\
\cref{ex:symp_corr} & $V_i(x)$ & (2 $\tanh$, 1 linear), 5 \\
\cref{ex:Cu_to_Pb} & $V_i(x)$ & (2 $\tanh$, 1 linear), 5\\
\cref{ex:analytic} & $V_i(x)$ & (2 $\tanh$, 1 linear), 5\\
\makecell[c]{\cref{sec:hydrocode}\\(AGC)}& $\hat{f}_G$& (4 $\tanh$, 1 linear), 20 \\
\makecell[c]{\cref{sec:hydrocode}\\(SC, ESC)}&$V_i(x)$& (2 $\tanh$, 1 linear), 20\\
\cref{exmp:toy} & $V_i(x)$ & (2 $\tanh$, 1 linear), 5\\
\bottomrule
\end{tabular}
\caption{Architecture specifications for each numerical example. Note that for $\hat{f}_G,\hat{f}_G^+$ the entire correction function is estimated by a fully connected neural network, while for the remaining examples with symplectic networks, only each component $V_i$ of each H\'{e}non map (\cref{def:henon_map}) is estimated by a fully connected network.}
\label{tbl:architectures}
\end{table}
\section{Hydrodynamic Simulation Details}\label{app:Hydrodynamic_Details}

To examine the ability of the machined learned EoS Tables to perform within the confines of a hydrodynamic simulation, a hydrodynamic test problem was constructed using the Los Alamos computational fluid dynamics program  in which a series of Tantalum Shells with initial velocities given as depicted in \cref{fig:InitialSystem} explodes into a Lead and Copper System.  

Pagosa is a 3D finite difference/volume Eulerian hydrodynamics program utilized for the study of high-pressure and high rate-rate deformation. .\cite{subramanian2020pagosa}   A volume fluid equation is incorporated, due to the multi-material problems of interest.  To close the the system of equations a EoS along with stress strain relationships are utilized.  In this investigation we utilize Steinberg strength models for both Copper and Lead. .\cite{steinberg1989constitutive,steinberg1980constitutive} 

Additional details of Pagosa include a Youngs Material Interface reconstruction algorithm to enable accurate shock capturing, a second order operator splitting algorithm, artificial viscosity, and a second order predictor corrector time integration methods.\cite{subramanian2020pagosa}

\end{document}